\documentclass[aps,prb,twocolumn,floats,floatfix,showpacs,footinbib]{revtex4}
\pdfoutput=1
\usepackage{calc}
\usepackage{tikz}
\usepackage{graphicx}
\usepackage{color}
\usepackage{amsmath, amssymb}
\usepackage[percent]{overpic}
\usepackage{array}
\usepackage{hyperref}
\usepackage{multirow}
\usepackage{comment}

\usetikzlibrary{arrows,calc}
\newcommand{\abs}[1]{\left| #1 \right|} 
\newcommand{\avg}[1]{\left< #1 \right>} 

\begin{document}
\title{Quantum percolation transition in 3d: density of states, finite size scaling and multifractality}
\author{L{\'a}szl{\'o} \surname{Ujfalusi}}
\author{Imre \surname{Varga}}
\email[Contact: ]{ujfalusi@phy.bme.hu, varga@phy.bme.hu}
\affiliation{Elm{\'e}leti Fizika Tansz{\'e}k, Fizikai Int{\'e}zet, 
Budapesti M{\H{u}}szaki {\'e}s Gazdas{\'a}gtudom{\'a}nyi Egyetem, 
H-1521 Budapest, Hungary}
\date{\today}
\begin{abstract}
The phase diagram of the metal-insulator transition in a three dimensional quantum percolation problem
is investigated numerically based on the multifractal analysis of the eigenstates. The large scale numerical simulation 
has been performed on systems with linear sizes up to $L=140$. The multifractal dimensions, exponents 
$D_q$ and $\alpha_q$, have been determined in the range of $0\leq q\leq 1$.
Our results confirm that this problem belongs to the same universality class as the three dimensional Anderson model, the
critical exponent of the localization length was found to be $\nu=1.622\pm 0.035$. 
However, the mulifractal function, $f(\alpha)$, and the exponents $D_q$ and $\alpha_q$ 
produced anomalous variations along the phase boundary, $p_c^Q(E)$.
\end{abstract}
\pacs{71.23.An,		
          71.30.+h,		
          72.15.Rn		
}
\maketitle


\section{Introduction}
The disorder induced metal-insulator transition, a genuine quantum phase transition is one of the most 
studied phenomena of condensed matter physics since the seminal paper published over five 
decades ago.~\cite{Anderson} According to the original problem, the Hamiltonian
\begin{equation} 
	\mathcal{H}=
		\sum_i \varepsilon_i a_i^\dagger a_i 
	     -t \sum_{\left<i,j\right>} \left( a_i^\dagger a_j + a_j^\dagger a_i \right),
\label{eq:am}
\end{equation}
describes the behavior of non-interacting spinless electrons in disorder. The first term in Eq.~(\ref{eq:am}) represents an
onsite disordered potential, where the energies, $\varepsilon_i$, are independent, uncorrelated random variables, drawn from a
distribution function, $P(\varepsilon)$, whose form is usually chosen to be uniform over an energy range that is symmetric 
around $\varepsilon=0$, but other forms, e.g. Gaussian or binary distributions could be used, as well. The second 
term in Eq.~(\ref{eq:am}) is the kinetic energy describing the hopping of the particles over a regular lattice, but restricted 
to nearest neighbors only. The energy scale associated to the hopping process, $t$, can be taken as the unit of energy ($t=1$).
The sites form a regular, usually simple cubic lattice. The embedding dimension, $d$, of the system is a very important
parameter, since phase transition occurs for $d>2$ only.~\cite{EversMirlin}

Besides diagonal disorder resembling substitutional disorder the other main cause of irregularity in condensed systems is
structural disorder. For the investigation of topological and structural disorder percolation is one of the most
important and widely used models. Percolation in general has a wide applicability in many fields of physics.~\cite{Stauffert}
In the Bernoulli site-percolation problem every site is filled with probability $p$ and is empty with probability $1-p$ independently. 
The main goal of classical percolation is to tell for a given $p$ whether an infinite cluster of filled sites may exist in the thermodynamical 
limit or not. It turns out, that there is such a critical probability, $p_c^{\scriptscriptstyle C}$, below which, $p<p_c^{\scriptscriptstyle C}$, 
there is no infinite cluster but above which, $p>p_c^{\scriptscriptstyle C}$, there is. In one dimension~\cite{perc1D} 
$p_c^{\scriptscriptstyle C}=1$, in two dimensions~\cite{perc2D} $p_c^{\scriptscriptstyle C}=0.592746216\pm0.00000013$, 
in three dimensions~\cite{Stauffert} $p_c^{\scriptscriptstyle C}=0.3116\pm0.0002$. In the $p>p_c^{\scriptscriptstyle C}$ case 
the existence of an infinite cluster ensures that the system can be treated as a conductor, since classical particles can travel 
through the whole system. On the other hand if  $p<p_c^{\scriptscriptstyle C}$, the system consists of a set of disjoint, 
finite clusters, and as a consequence, it behaves as an insulator, since no particle can escape from its initial finite cluster.

For the electric conduction properties of a sample the electrons are responsible whose behavior is described very well 
by quantum mechanics,  therefore we shall investigate spinless non--interacting electrons on a percolated lattice, 
this is called the quantum percolation model. Omitting spin and interaction is necessary, because even with these 
simplifications the problem seems to be hard to solve. The corresponding Hamiltonian is
\begin{equation} 
	\mathcal{H}=
		\sum_{i\in A} \varepsilon a_i^\dagger a_i 
		-\sum_{\substack{
			\left<i,j\right> \\
			i,j\in A}} 
		\left( a_i^\dagger a_j + a_j^\dagger a_i \right),
\label{eq:qperc_qperchami}
\end{equation}
where $A$ is the set of filled sites, $\varepsilon$ is a constant on--site energy, whose value can be safely set to zero without 
loss of generality. Note that the pure site--percolation problem is equivalent to a binary Anderson 
model~\cite{Kirkpatrick-Eggarter,Kusy,Soukoulis} with constant $\varepsilon_A$ and $\varepsilon_B$ but taking the limit 
$\varepsilon_B \to \infty$:
\begin{equation} 
\mathcal{H}=\sum_{i\in A} \varepsilon_A a_i^\dagger a_i + \sum_{i\in B} \varepsilon_B a_i^\dagger a_i 
            - \sum_{\left<i,j\right>} \left( a_i^\dagger a_j + a_j^\dagger a_i \right)
\end{equation}
This Hamiltonian could describe an alloy of a perfect metal consisting of atoms $A$ and a perfect insulator consisting of 
atoms $B$ only. All $A$ sites are equivalent, and the $B$ sites cannot be reached due to their infinite on-site energy, 
therefore $B$ sites behave as if they were empty. This suggests, that quantum percolation behaves similar to the Anderson model. 
In our present work we shall show many similarities. The most important similarity with the Anderson problem is
the existence of a metal--insulator transition for the quantum percolation model too, however, here $p$, or strictly 
speaking $(1-p)$, plays the role of disorder: For $p<p_c^{\scriptscriptstyle C}$ every state is localized onto finite, connected 
islands, thus the sample is an insulator. Increasing $p$ beyond $p_c^{\scriptscriptstyle C}$, however, a classical particle 
can travel through the sample, the electron wave functions are localized due to strong interference effects caused by disorder, 
the sample still remains an insulator. For $p$ values slightly below $1$ states are perturbed Bloch-states, the sample is a metal. 
In between there exists a mobility edge, $p_c^{\scriptscriptstyle Q}(E)$, an energy--dependent quantum critical point, 
below which electronic eigenstates are Anderson-localized giving rise to an insulator, and above which they are extended forming 
a metal. Along the mobility edge, $p_c^{\scriptscriptstyle Q}(E)$, the states are supposed to be multifractals. 
In Sec.~\ref{sec:fss_qperc} we argue, that the Anderson model and the quantum percolation model belong to the same 
universality class. 

The organization of the paper is the following. In the next section, Sec.~\ref{sec:theory} we look at the peculiar properties 
of the density of states in quantum percolation and provide an overview about multifractality together with an introduction
about the finite-size scaling analysis of the corresponding generalized dimensions. In Sec.~\ref{subsec:fss_anderson} 
we give a short overview of the technique of the latter analysis in the case of the 3D Anderson transition, 
in Sec.~\ref{sec:fss_qperc} we provide with the methods applied in the present work, and in Sect.~\ref{sec:gmfe_qperc}
we present the results of our analysis for the multifractal analysis. Finally Sec.~\ref{sec:summ} is left for a summary.

\section{Theoretical and numerical background}
\label{sec:theory}
Electronic conduction is only possible on an infinite cluster, so $p_c^{\scriptscriptstyle Q} > p_c^{\scriptscriptstyle C}$ is expected, 
therefore the infinite cluster should be investigated, so only the $p>p_c^{\scriptscriptstyle C}$ regime is interesting for us.
Since numerically we can deal with a finite lattice only, we restricted our work on 
the largest finite cluster found by a Hoshen-Kopelman algorithm~\cite{Hoshen-Kopelman}. In a finite size sample the 
Hamiltonian, Eq.~(\ref{eq:qperc_qperchami}) is a huge sparse matrix. To obtain the spectrum and eigenfunctions we used the 
Jacobi-Davidson method encoded in the PRIMME package~\cite{Stathopoulos10} with ILU preconditioning, using the ILUPACK 
package~\cite{Bollhofer08}. 

At first let us take a glance at the density of states (DOS) because for the quantum percolation problem it deserves a 
special attention.

\subsection{Density of states}
\label{sec:qperc_dos}
\begin{figure*}
	\begin {center}
	\begin{tabular}{c c c}
	\begin{overpic}[type=pdf,ext=.pdf,read=.pdf,width=.33\linewidth]{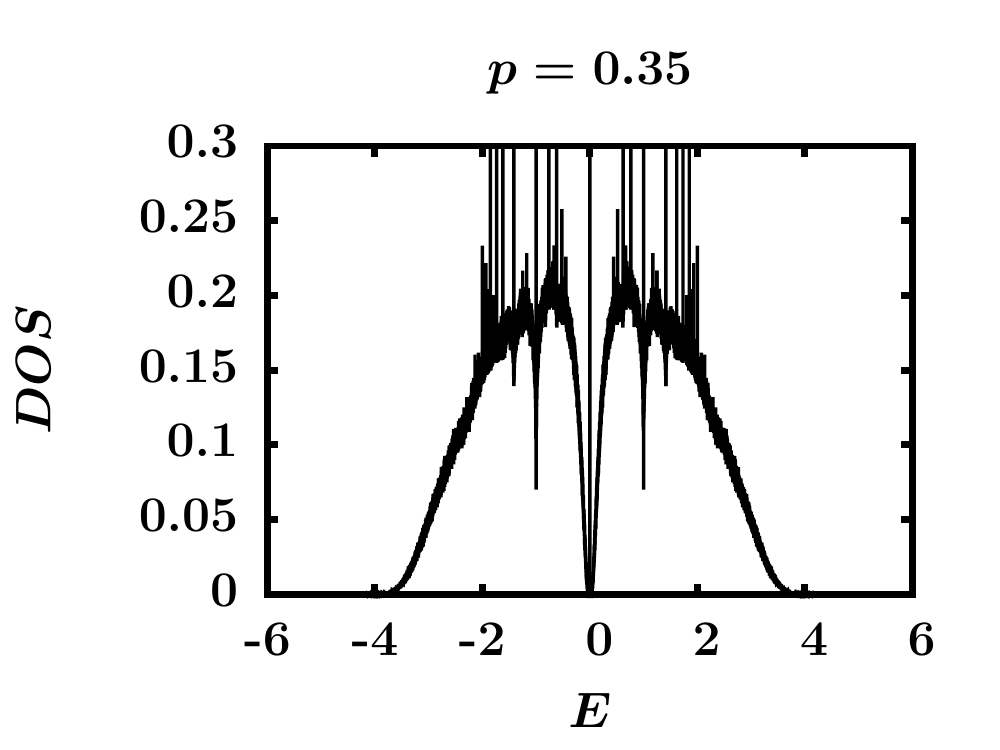} \put(0,70){(a)} \end{overpic} & 
	\begin{overpic}[type=pdf,ext=.pdf,read=.pdf,width=.33\linewidth]{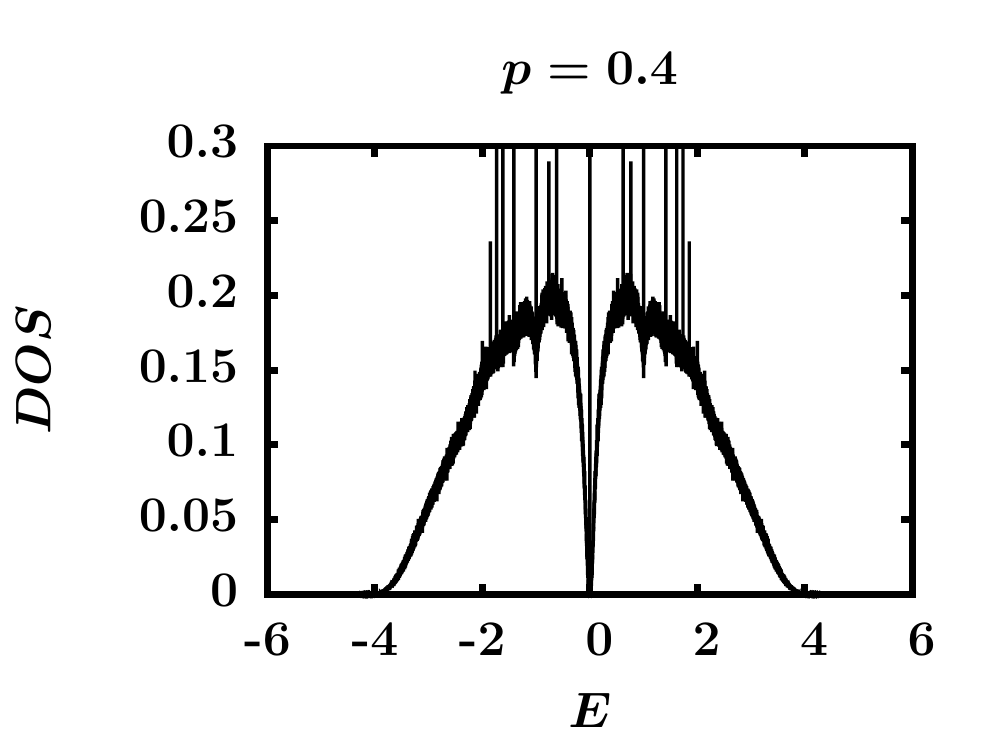} \put(0,70){(b)} \end{overpic} &
	\multirow{2}[2]{*}[3.5cm]{\begin{overpic}[type=png,ext=.png,read=.png,width=.33\linewidth]{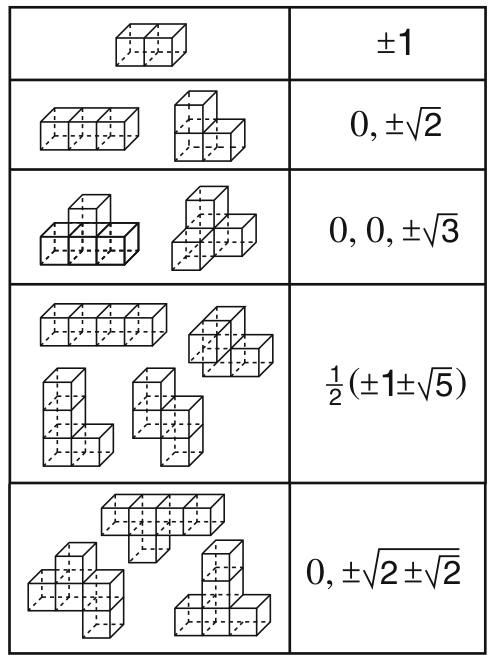} \put(0,103){(e)} \end{overpic}} \\
	\begin{overpic}[type=pdf,ext=.pdf,read=.pdf,width=.33\linewidth]{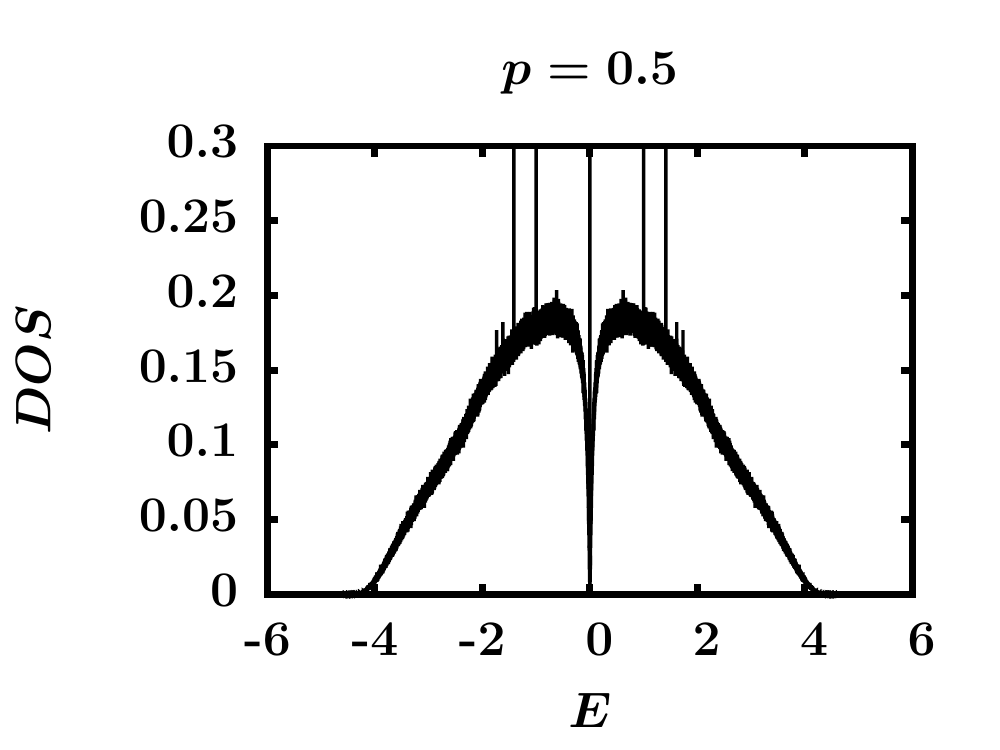}  \put(0,70){(c)} \end{overpic}&
	\begin{overpic}[type=pdf,ext=.pdf,read=.pdf,width=.33\linewidth]{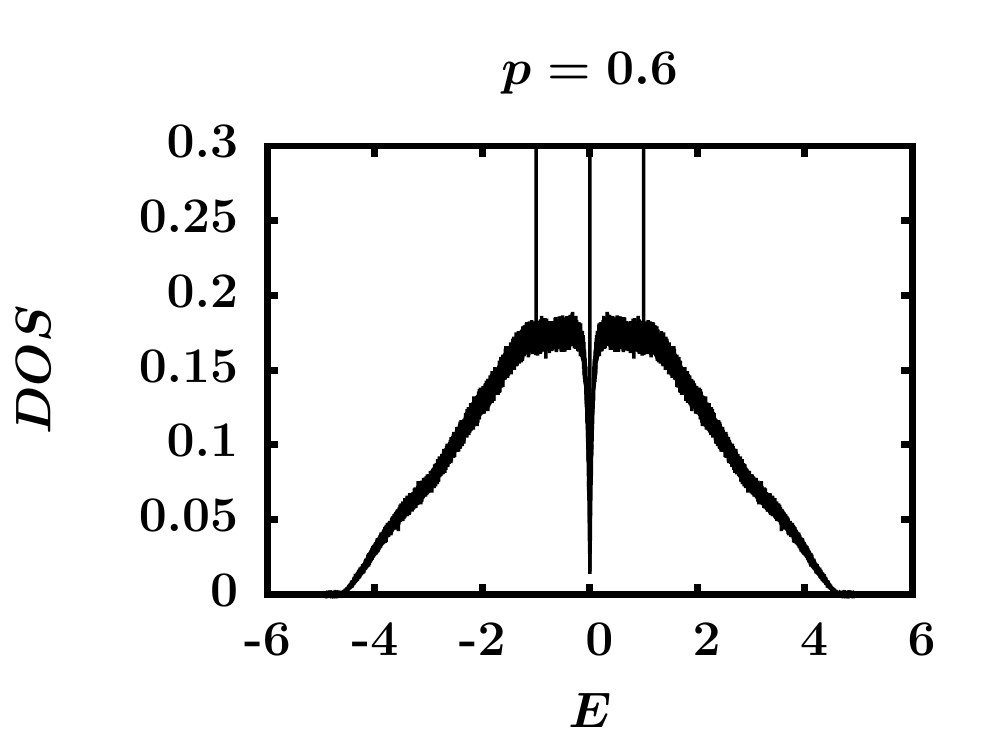}  \put(0,70){(d)} \end{overpic}& \\ 
	\end{tabular}
	\end{center}
	\caption{Left side: Density of states of quantum percolation model at different site-filling probabilities, (a) $p=0.35$, (b) $p=0.4$, (c) $p=0.5$,
	(d) $p=0.6$. Panel (e):  Small clusters corresponding to special energies taken from the review of Schubert and Fehske~\cite{Fehske-Schubert}.} 
	\label{fig:qperc_DOS}	
\end{figure*}

The DOS of the giant cluster has itself an unusual form. The evolution of this function with $p$ is depicted 
in Fig.~\ref{fig:qperc_DOS}. With increasing disorder, in the present case this means decreasing $p$, more and more sharp 
peaks appear in the spectrum. These peaks correspond to special, so-called ''molecular states'', which are localized to a
few sites~\cite{Kirkpatrick-Eggarter}. These states are non-zero on a few sites only and exactly zero on every other one 
due to exact destructive interference. Therefore they are not localized in the sense of Anderson localization, 
there is no exponential decay in the wave function envelope. Typical few-site structures and corresponding energies are given on 
the right side of Fig.~\ref{fig:qperc_DOS}. Since the value $E=0$ appears for most clusters as an eigenvalue, 
the highest peak of the DOS is at the middle of the band, and there is also a pseudo--gap around it. 
Considering other few-site clusters there is no reason for the eigenvalues to avoid any part of the band, therefore peaks in the 
DOS corresponding to molecular states should appear densely in the thermodynamic limit. The energy of a molecular state is a 
strict value, thus the peaks in the DOS appear as a series of Dirac-deltas. As we can see, the spectrum consists of 
two parts: a dense point spectrum due to molecular states, and a continuous one due to all other states.~\cite{Kirkpatrick-Eggarter} 
This statement has been rigorously proven recently in the case of a 2D square lattice, and for tree graphs 
corresponding to an effective infinite dimension, therefore it is conjectured to be true in any dimension.~\cite{Virag14}

Since molecular states are strongly localized, they cannot contribute to conduction. Therefore we restrict our investigation to 
the continuous part of the spectrum only. With the numerical method described above we are able to compute one single eigenstate 
of the Hamiltonian having an eigenenergy close to a given value of $E$. In Fig.~\ref{fig:qperc_DOS} it is shown, 
that in a finite system molecular states appear frequently at few special energies only, e.g. $E=0, \pm 1, \pm\sqrt{2} \dots$, 
therefore for our purpose we have chosen energy windows avoiding the peaks in the DOS. 

The cubic lattice is a bipartite lattice and the Hamiltonian (\ref{eq:qperc_qperchami}) couples nearest neighbors only, 
therefore from one sublattice, $\alpha$, it is possible to hop to the other sublattice, $\beta$, only. The Hamiltonian anticommutes with an
operator $C$, which is $1$ on sublattice $\alpha$, and $-1$ on sublattice $\beta$, thus $C$ acts as a chirality transformation.~\cite{Naumis02} 
Therefore the quantum percolation model is symmetric not only on average for the exchange of eigenenergies, $-E\leftrightarrow E$, but for every 
single disorder realization. In the low (high) energy range the states have antibonding (bonding) character.  
In the middle of the band, around $E=0$, chessboard-like chiral states appear. These chiral states exactly at $E=0$ are eigenfunctions of $C$, 
as well, therefore they are protected against off-diagonal disorder.

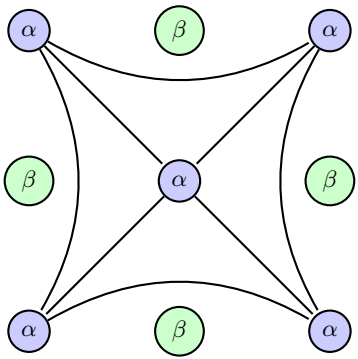
\begin{figure}
\begin{center}
\begin{tikzpicture}[-,>=stealth',shorten >=1pt,auto,node distance=2cm,
  thick,A node/.style={circle,fill=blue!20,draw},B node/.style={circle,fill=green!20,draw}]

  \node[A node] (1) {$\alpha$};
  \node[B node] (2) [right of=1] {$\beta$};
  \node[A node] (3) [right of=2] {$\alpha$};
  \node[B node] (4) [below of=1] {$\beta$};
  \node[A node] (5) [below of=2] {$\alpha$};
  \node[B node] (6) [below of=3] {$\beta$};
  \node[A node] (7) [below of=4] {$\alpha$};
  \node[B node] (8) [below of=5] {$\beta$};
  \node[A node] (9) [below of=6] {$\alpha$};

  \path[every node/.style={font=\sffamily\small}]
    (1) edge [bend right] (3)
        edge [] (5)
        edge [bend left] (7)
    (3) edge [] (5)
        edge [bend right] (9)
    (5) edge [] (7)
        edge [] (9)
    (7) edge [bend left] (9);
\end{tikzpicture}
\caption{Hopping elements in the ,,renormalized'' Hamiltonian, $\mathcal{H}^2$}
\label{fig:qperc_renorm_Hami}
\end{center}
\end{figure}

In order to understand the sub gap appearing around the middle of the band, $E=0$, we invoke the arguments of Ref.~\onlinecite{Naumis02}.
The square of the Hamiltonian, $\mathcal{H}^2$, connects the sites of the same sublattice only, see Fig.~\ref{fig:qperc_renorm_Hami}, 
thus one can ,,renormalize'' $\mathcal{H}^2$ acting on one of the sublattices~\cite{Naumis02}. 
The vicinity of $E=0$ belongs to the low-energy regime of the spectrum of $\mathcal{H}^2$, therefore here antibonding states should appear, 
which are more or less visible in the wave functions themselves, too. But the hopping elements to the diagonal-lying second neighbors 
in Fig.~\ref{fig:qperc_renorm_Hami} introduce triangles. Triangles and the antibonding nature together lead to frustration. 
Based on the frustration of the states around zero energy Naumis {\it et. al}~\cite{Naumis02} showed in two dimensions, 
that the width of the pseudogap around zero energy, $\Delta$, is connected to the peak at $E=0$: $\Delta\sim \sqrt{\rho_0}$, 
where $\rho_0$ stands for the weight of the zero energy states in the spectra. They also showed, that the width of the pseudogap tends 
to zero in the non-disordered limit, $\lim_{p\to 1}\Delta=0$. The extension of these arguments to three dimensions should be valid, 
since the most important ingredient of their calculation is the coordination number of the lattice, and not the dimensionality itself explicitly.

The states close to $E=0$ belong to the edge of the spectrum of $\mathcal{H}^2$, which is a disordered Hamiltonian. 
Therefore the pseudogap might be qualitatively interpreted as the Lifshitz tail of $\mathcal{H}^2$, leading to localized states close $E=0$.

\subsection{Introduction to multifractals}
\label{sec:multifractals}
\begin{figure*}
	\begin {center}
	\begin{tabular}{c c c}
	metal/extended & critical/multifractal & insulator/localized\\ \\
	\begin{overpic}[type=pdf,ext=.png,read=.png,width=.33\linewidth]{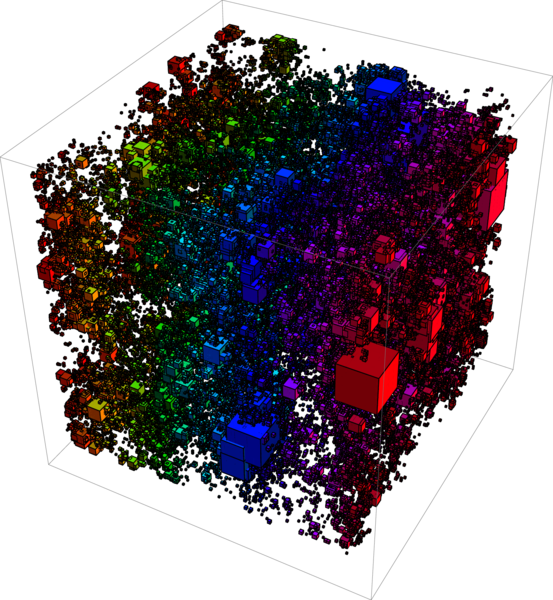} \put(0,100){(a)} \end{overpic} & 
	\begin{overpic}[type=pdf,ext=.png,read=.png,width=.33\linewidth]{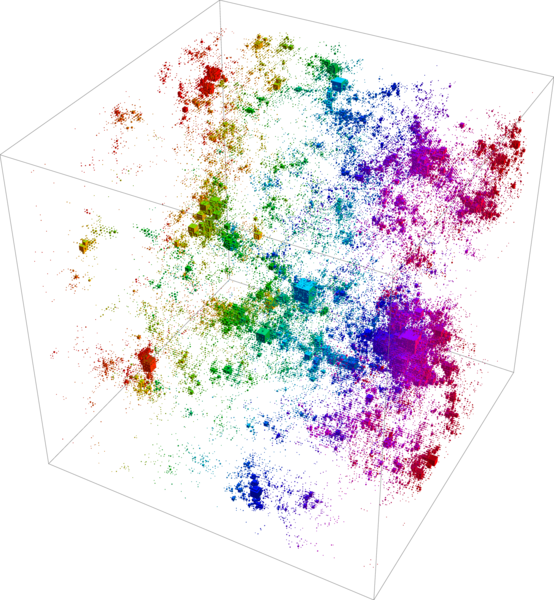} \put(0,100){(b)} \end{overpic} & 
	\begin{overpic}[type=pdf,ext=.png,read=.png,width=.33\linewidth]{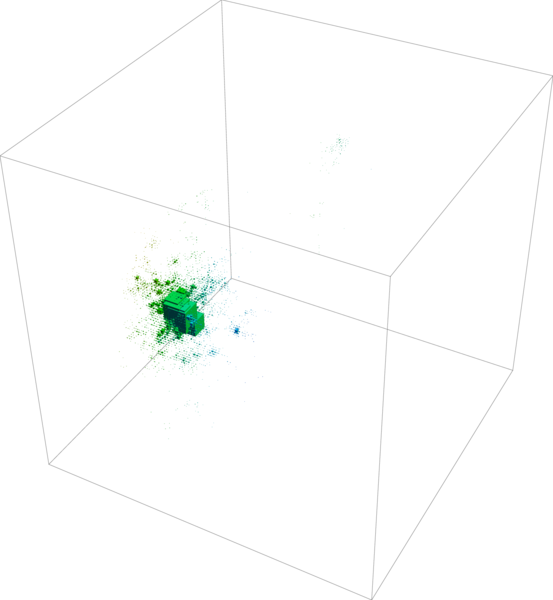} \put(0,100){(c)} \end{overpic} \\
	\begin{overpic}[type=pdf,ext=.png,read=.png,width=.33\linewidth]{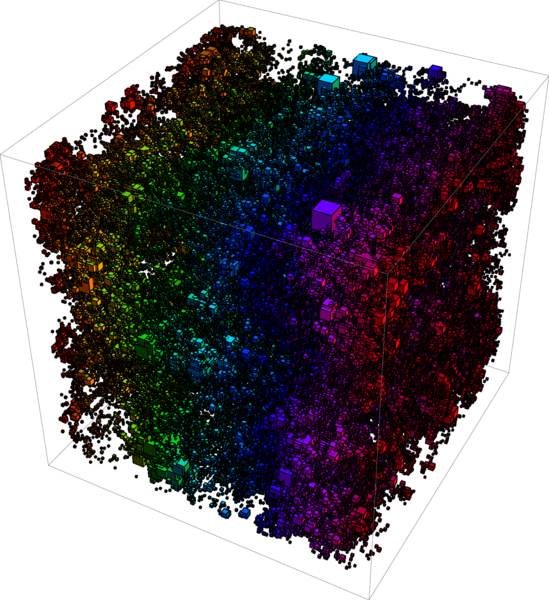} \put(0,100){(d)} \end{overpic} & 
	\begin{overpic}[type=pdf,ext=.png,read=.png,width=.33\linewidth]{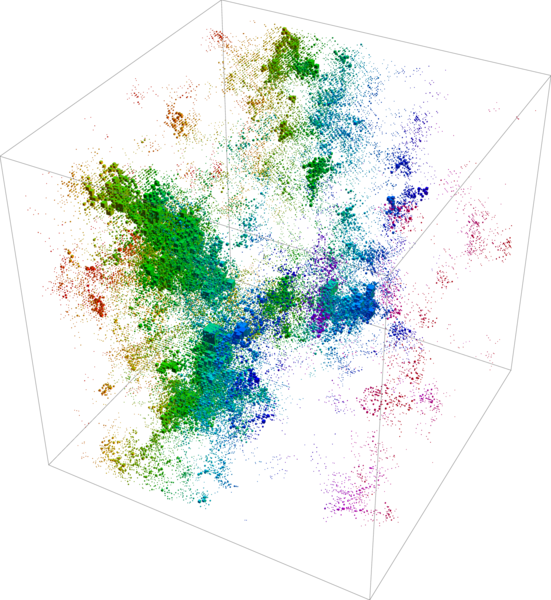} \put(0,100){(e)} \end{overpic} & 
	\begin{overpic}[type=pdf,ext=.png,read=.png,width=.33\linewidth]{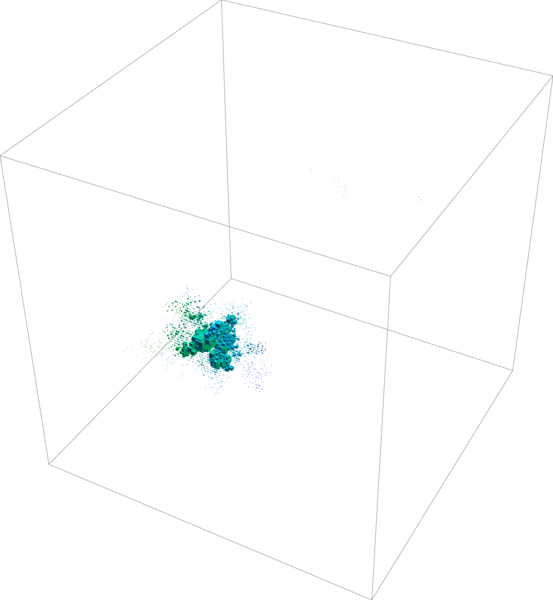} \put(0,100){(f)} \end{overpic} \\
	\end{tabular}
	\end{center}
	\caption{First row: Eigenvectors of the Anderson-model at $E=0$ (a) on the metallic side at $W=14$, (b) close to criticality $W=16.5$ and 
	(c) on the insulating side at $W=20$. Second row: Eigenvectors of the Quantum percolation model at energy $E=0.1$ (d) on the metallic side at $p=0.5$, 
	(e) close to criticality $p=0.4535$ and (f) on the insulating side at $p=0.4$. Box sizes correspond to (a),(d) $400\cdot\sqrt{|\Psi|^2}$ (b), 
	(e) $70\cdot\sqrt{|\Psi|^2}$ (c),(f) $20\cdot\sqrt{|\Psi|^2}$. Multiplying factors were tuned to best sight but without overlapping cubes. 
	System size, $L=120$, for all subfigures. Coloring is due to $x$ coordinate.}
	\label{fig:fss_eigvec_anderson_qperc}	
\end{figure*}

In recent high-precision calculations~\cite{Rodriguez11} the so-called Multifractal Exponents (MFE) have been 
used to describe the Anderson metal--insulator transition (AMIT). The renormalization flow of the AMIT as mentioned in the
Introduction has three fixed points: a metallic, an insulating and a critical one. In the metallic fixed point every state 
is extended with probability one, thus with increasing system size, the effective size of the states also grows proportional 
to the volume. So the fractal dimension of the states, that will be defined more precisely later, is just the embedding dimension $q$-independently, 
$D_q^{met}\equiv d$. In the insulating fixed point every state is exponentially localized, their effective size does not 
change with growing system size, thus for $q\geq 0$ $D_q^{ins}\equiv 0$, for $q< 0$ $D_q^{ins}\equiv \infty$. 
At criticality the system does not change during renormalization, thus it must be statistically the same on all length scales 
showing scale independence, which means self similarity. Therefore wave functions are multifractals, in other words generalized 
fractals~\cite{janssen}, see Fig.~\ref{fig:fss_eigvec_anderson_qperc}.
In our case there is a $d$-dimensional hypercubic lattice with linear size $L$, and a normalized wave function whose support
is this lattice, $\sum\limits_{i=1}^{L^{d}}\abs{\Psi_i}^2=1$, defining a probability distribution. Let us divide this lattice 
into smaller hypercubes (boxes) with linear size $\ell$, and introduce the ratio $\lambda=\frac{\ell}{L}$. 
Then coarse graining $|\psi|^2$, in other words summing all its values in the $k$th box we obtain:
\begin{equation} 
\mu_k=\sum_{i\in box_k} \abs{\Psi_i}^2,
\end{equation}
where $\mu_k$ is the weight associated to the $k$th box termed as box--probability. Let us define the $q$th moment of the 
mass, frequently called generalized inverse participation ratio (GIPR), and its derivative as
\begin{equation} 
R_q=\sum_{k=1}^{\lambda^{-d}} \mu_k^q=\lambda^{\tilde{\tau}_q}\qquad
S_q=\frac{dR_q}{dq}=\sum_{k=1}^{\lambda^{-d}} \mu_k^q \ln \mu_k,
\label{eq:SqRq}
\end{equation}
where $\tilde{\tau}_q$ is the finite system mass exponent. $\tilde{\tau}_q$ and its derivative read as:
\begin{equation} 
\tilde{\tau}_q=\frac{\ln R_q}{\ln \lambda}\qquad \tilde{\alpha}_q=\frac{d\tilde{\tau}_q}{dq}=\frac{S_q}{R_q\ln\lambda}.
\end{equation}
Taking the $L\to\infty$ limit, which is equivalent to taking the $\lambda\to 0$ limit, the mass exponent and its derivative are
\begin{equation} 
\tau_q=\lim_{\lambda\to 0}\frac{\ln R_q}{\ln \lambda} \qquad 
\alpha_q=\frac{d\tau_q}{dq}=\lim_{\lambda\to 0}\frac{S_q}{R_q\ln\lambda}.
\end{equation}
$\tau_q$ can be written in the form
\begin{equation} 
\tau_q=D_q(q-1)=d(q-1)+\Delta_q,\label{eq:multifractal_tauq_Dq_deltaq}
\end{equation}
where $D_q$ is the generalized fractal dimension. In this expression $\Delta_q$ is the anomalous scaling exponent:
\begin{equation} 
D_q=\frac{1}{q-1} \lim_{\lambda\to 0}\frac{\ln R_q}{\ln \lambda}\qquad \Delta_q=(D_q-d)(q-1).
\end{equation}
The quantities $\tau_q$, $\alpha_q$, $D_q$, and $\Delta_q$ are often referred as multifractal exponents (MFEs), while the finite system version of these exponents, $\tilde{\tau}_q,\ \tilde{\alpha}_q,\ \tilde{D}_q\text{ and }\tilde{\Delta}_q$, are called generalized multifractal exponents (GMFEs). 
$D_q$ is directly related to the so-called {\it R\'{e}nyi-entropy}, $H_q=(q-1)^{-1}\ln R_q$, which in the limit $q\to 1$
yields the well-known {\it Shannon-entropy}, i.e. $-\sum_k \mu_k \ln \mu_k$. This is the reason why $D_1$ is also referred 
as information dimension:
\begin{equation} 
D_1=\lim_{q\to 1} \frac{1}{q-1} \lim_{\lambda\to 0}\frac{\ln R_q}{\ln \lambda} \stackrel{L'H}{=} 
\alpha_1=\lim_{\lambda\to 0}\frac{1}{\ln\lambda} \sum\limits_{k=1}^{\lambda^{-d}} \mu_k \ln \mu_k,
\end{equation}
while another frequently used dimension is the correlation dimension, $D_2$. The latter dimension appeared often in
recent studies of the physical relevance of multifractal eigenstates~\cite{cuevas}

There is another way to characterize the multifractal nature of the wave functions. For that purpose the  box probability, 
$\mu$ can be transformed into another variable, $\alpha=\ln \mu/\ln \lambda$ assuming the fractal scaling
\begin{equation}
    \mu\sim \lambda^{\alpha}\qquad .
\end{equation}
Let us denote the probability density function of the number of boxes having a value $\alpha$ with $\mathcal{P}(\alpha)$. 
The scaling of $\mathcal{P}(\alpha)$ is described through the singularity spectrum $f(\alpha)$, which is the fractal dimension 
of the number of boxes having a value $\alpha$:
\begin{equation} 
    \mathcal{P}(\alpha)\sim\lambda^{f(\alpha)}
\label{eq:multifractal_falpha}
\end{equation}
Function $f(\alpha)$ is nothing else but the Legendre-transform of $\tau_q$:
\begin{equation} 
    f(\alpha_q)=q\alpha_q-\tau_q.
\label{eq:multifractal_falpha_tauq}
\end{equation}

According to recent results a symmetry relation exists for $\alpha_q$ and $\Delta_q$ given in the form~\cite{Mirlin06}:
\begin{equation} 
    \Delta_q-\Delta_{1-q}=0\qquad\qquad \alpha_q+\alpha_{1-q}=2d 
\label{eq:multifractals_Deltaalphasymmety}
\end{equation}
This relation first obtained for some random matrix ensemble numerically and using the supersymmetric non-linear 
sigma model analytically~\cite{Mirlin06} was later confirmed for several two dimensional~\cite{milden07,evers08} and
three dimensional systems~\cite{vasquez08}. However, deviations have been detected in other cases.~\cite{subra06,faez09}
The robustness of this relation has been investigated also for many-body localization.~\cite{monthus11}

%
%
\subsection{Finite size scaling laws for GMFEs}
\label{sec:MFSS}
Finite size scaling techniques are very well described by Rodriguez {\it et. al}~\cite{Rodriguez11} for the Anderson model. 
We are going to use their notation, therefore we denote disorder by $W$. In this subsection we extend the formalism of
Ref.~\onlinecite{Rodriguez11}. From the eigenfunction the $R_q$ and 
$S_q$ values can be computed for every state at different $q$ values. At fixed disorder, $W$, system size, $L$, and box size, 
$\ell$, every GMFE is computable from these two quantities the following way~\cite{Rodriguez11}:
\begin{widetext}
\begin{subequations}
\begin{align}
\tilde{\tau}_q^{ens}(W,L,\ell) &=\frac{\ln\avg{R_q}}{\ln\lambda} &\qquad& 
\tilde{\tau}_q^{typ}(W,L,\ell)=\frac{\avg{\ln R_q}}{\ln\lambda}\label{eq:fss_FSGMFE_sub1}\\
\tilde{\alpha}_q^{ens}(W,L,\ell)&=\frac{\avg{S_q}}{\avg{R_q}\ln\lambda} &\qquad& 
\tilde{\alpha}_q^{typ}(W,L,\ell)=\avg{\frac{S_q}{R_q}}\frac{1}{\ln\lambda}\label{eq:fss_FSGMFE_sub2}\\
\tilde{D}_q^{ens}(W,L,\ell)&=\frac{1}{q-1}\frac{\ln\avg{R_q}}{\ln\lambda} &\qquad& 
\tilde{D}_q^{typ}(W,L,\ell)=\frac{1}{q-1}\frac{\avg{\ln R_q}}{\ln\lambda}\label{eq:fss_FSGMFE_sub3}\\
\tilde{\Delta}_q^{ens}(W,L,\ell)&=\frac{\ln\avg{R_q}}{\ln\lambda}-d(q-1) &\qquad& 
\tilde{\Delta}_q^{typ}(W,L,\ell)=\frac{\avg{\ln R_q}}{\ln\lambda}-d(q-1),\label{eq:fss_FSGMFE_sub4}
\end{align}
\end{subequations}
\end{widetext}
where $\left<.\right>$ stands for averaging: $ens$ and $typ$ denote the {\it ensemble} and {\it typical} averaging. 
Every GMFE approach the value of the corresponding MFE at the critical point only in the limit $\lambda\to 0$.
Close to the critical point due to standard finite size scaling arguments we can suppose, that $R_q$ and $S_q$ shows 
scaling behavior determined only by the ratio of two length scales, $L$ and $\ell$, and the localization/correlation 
length, ${\xi}$, in the insulating/metallic phase:
\begin{equation}
R_q(W,L,\ell)=\lambda^{\tau_q}\mathcal{R}_q\left(\frac{L}{\xi},\frac{\ell}{\xi}\right)
\label{eq:fss_RWLl}\end{equation}
According to (\ref{eq:fss_FSGMFE_sub1})--(\ref{eq:fss_FSGMFE_sub4}) for all GMFEs the scaling-law holds independently 
from the type of averaging~\cite{Rodriguez11}:
\begin{subequations}
\begin{align}
\tilde{\tau}_q(W,L,\ell) &= \tau_q+\frac{q(q-1)}{\ln\lambda}\mathcal{T}_q\left(\frac{L}{\xi},\frac{\ell}{\xi}\right)\label{eq:fss_tauWLl}\\
\tilde{\alpha}_q(W,L,\ell) &= \alpha_q+\frac{1}{\ln\lambda}\mathcal{A}_q\left(\frac{L}{\xi},\frac{\ell}{\xi}\right)
\label{eq:fss_alphaWLl}\\
\tilde{D}_q(W,L,\ell) &= D_q+\frac{q}{\ln\lambda}\mathcal{T}_q\left(\frac{L}{\xi},\frac{\ell}{\xi}\right)
\label{eq:fss_DWLl}\\
\tilde{\Delta}_q(W,L,\ell) &= \Delta_q+\frac{q(q-1)}{\ln\lambda}\mathcal{T}_q\left(\frac{L}{\xi},\frac{\ell}{\xi}\right)\label{eq:fss_deltaWLl},
\end{align}
\end{subequations}
Equations (\ref{eq:fss_tauWLl})--(\ref{eq:fss_deltaWLl}) can be summarized in one equation:
\begin{equation} \tilde{G}_q(W,L,\ell) = G_q+\frac{1}{\ln\lambda}\mathcal{G}_q\left(\frac{L}{\xi},\frac{\ell}{\xi}\right)\label{eq:fss_scalinglaw_Ll}\end{equation}
$(L,\ell)$ on the left and $\left(\frac{L}{\xi},\frac{\ell}{\xi}\right)$ on the right hand side can be changed to $(L,\lambda)$ and $\left(\frac{L}{\xi},\lambda\right)$:
\begin{equation} \tilde{G}_q(W,L,\lambda) = G_q+\frac{1}{\ln\lambda}\mathcal{G}_q\left(\frac{L}{\xi},\lambda\right)\label{eq:fss_scalinglaw_Llambda}
\end{equation}

\subsubsection{Finite size scaling at fixed $\lambda$}
\label{fss_fixed_lambda}
At fixed $\lambda$, $G_q$ in Eq.~(\ref{eq:fss_scalinglaw_Llambda}) can be considered as the constant term of $\mathcal{G}_q$, 
therefore
\begin{equation} 
     \tilde{G}_q(W,L) = \mathcal{G}_q\left(\frac{L}{\xi}\right),
\label{eq:fss_anderson_scalinglaw}
\end{equation}
where the constant $\lambda$ has been dropped. $\mathcal{G}_q$ can be expanded with one relevant, $\varrho(w)$, 
and one irrelevant operator, $\eta(w)$, the following way using $w=W-W_c$:
\begin{equation} 
     \mathcal{G}_q\left(\varrho L^{\frac{1}{\nu}}, \eta L^{-y}\right) = 
\mathcal{G}^{rel}_q\left(\varrho L^{\frac{1}{\nu}}\right) + \eta L^{-y}\mathcal{G}^{irrel}_q\left(\varrho L^{\frac{1}{\nu}}\right)
\end{equation}
All the disorder-dependent quantities in the above formula can be expanded in Taylor-series:
\begin{eqnarray} 
     \mathcal{G}^{rel}_q\left(\varrho L^{\frac{1}{\nu}}\right)&=&\sum_{i=0}^{n_{rel}} a_i\left(\varrho L^{\frac{1}{\nu}}\right)^i\\
     \mathcal{G}^{irrel}_q\left(\varrho L^{\frac{1}{\nu}}\right)&=&\sum_{i=0}^{n_{irrel}} b_i\left(\varrho L^{\frac{1}{\nu}}\right)^i\\
     \varrho(w)=w+\sum_{i=2}^{n_{\varrho}}c_i w^i &\quad& \eta(w)=1+\sum_{i=1}^{n_{\eta}}d_i w^i
\end{eqnarray}
The number of parameters is $n_{rel}+n_{irrel}+n_{\rho}+n_{\eta}+1$.

\subsubsection{Finite size scaling at fixed $\ell=1$}
\label{fss_fixed_l}
For fixed $\ell$ the scaling law given in Eq.~(\ref{eq:fss_scalinglaw_Ll}) has to be considered. The expansion of $\mathcal{G}$ 
in (\ref{eq:fss_scalinglaw_Ll}) is
\begin{eqnarray*} & &\mathcal{G}_q\left(\varrho L^{\frac{1}{\nu}},\varrho \ell^{\frac{1}{\nu}}, \eta L^{-y}, 
\eta' \ell^{-y'}\right) = \mathcal{G}^{rel}_q\left(\varrho L^{\frac{1}{\nu}}, \varrho \ell^{\frac{1}{\nu}}\right) +\\
 & &+\eta L^{-y}\mathcal{G}^{irrel}_q\left(\varrho L^{\frac{1}{\nu}},\varrho \ell^{\frac{1}{\nu}}\right) +
\eta' \ell^{-y'}\mathcal{G'}^{irrel}_q\left(\varrho L^{\frac{1}{\nu}},\varrho \ell^{\frac{1}{\nu}}\right).
\end{eqnarray*}
Choosing $\ell=1$, and considering that in most cases $\eta$ and $\eta'$ are constant, i.e. $n_{\eta}=0$, the last term can be merged with the relevant part. 
Equation (\ref{eq:fss_scalinglaw_Ll}) has the following form for fixed $\ell=1$:
\begin{eqnarray}
\label{eq:fss_anderson_Gammal}& &\tilde{G}_q(W,L)=\\
& &G_q + \frac{1}{\ln L}\left(\mathcal{G}^{rel}_q\left(\varrho L^{\frac{1}{\nu}}, \varrho\right) + 
\eta L^{-y}\mathcal{G}^{irrel}_q\left(\varrho L^{\frac{1}{\nu}}, \varrho\right) \right).
\nonumber
\end{eqnarray}
The Taylor-expansions of the above functions are
\begin{eqnarray} 
\mathcal{G}^{rel}_q\left(\varrho L^{\frac{1}{\nu}},\varrho \right) &=& \sum_{i=0}^{n_{rel}}\sum_{j=0}^{i} a_{ij}\varrho^i L^{\frac{j}{\nu}}\\
\mathcal{G}^{irrel}_q\left(\varrho L^{\frac{1}{\nu}},\varrho\right) 
&=&\sum_{i=0}^{n_{rel}}\sum_{j=0}^{i} b_{ij}\varrho^i L^{\frac{j}{\nu}} \label{eq:fss_anderson_Grelirrel}\\
\varrho_{(w)}=w+\sum_{i=2}^{n_{\varrho}}c_i w^i\ & & \ \eta_{(w)}=1+\sum_{i=1}^{n_{\eta}}d_i w^i
\end{eqnarray}
The number of parameters is $3n_{rel}(n_{rel}+1)/2+3n_{irrel}(n_{irrel}+1)/2+n_{\rho}+n_{\eta}-1$. We can see, that the number of parameters grows 
as $\sim n_{rel/irrel}^2$ for fixed $\ell=1$, instead of $\sim n_{rel/irrel}$ as for fixed $\lambda$. This makes the fitting procedure definitely much more 
difficult.

\section{Finite size scaling for the 3D quantum percolation model using GMFEs} 
\label{sec:fss_qperc}
Before turning to the analysis of our simulations on the 3D quantum percolation model, we briefly review the details of the finite size scaling using
GMFEs but first based on the 3D Anderson model. The aim of this section is twofold. First of all we present the advantages and disadvantages 
of the various methods used and their applicability for our purposes. Second we show the precision of these techniques for the case of a well-studied
case, the Anderson transition.
\subsection{Finite size scaling for the 3D Anderson model using GMFEs}
\label{subsec:fss_anderson}
Our first goal was to check our numerical algorithm on the well-known Anderson problem. Based on Ref.~\onlinecite{Rodriguez11} we formulate
two cases: at first fixing $\lambda$ then fixing $\ell$.
\subsubsection{Finite size scaling at fixed $\lambda$}
Since the metal-insulator transition occurs at the band center~\cite{EversMirlin} ($E=0$) at disorder $W_c\approx 16.5$, most works study the vicinity of this point. To have the best comparison, we analyzed this regime also, therefore about $20$ disorder 
values were taken for the range $15\leq W \leq 18$. System sizes were taken from the range $L=20..100$, the number of 
samples were $N=4000$ at least. We considered only one wave function per realization, the one with energy closest to 
zero in order to avoid correlations between wave functions of the same system~\cite{Rodriguez11}. From the wave function 
the $R_q$ and $S_q$ multifractal moments were calculated in the range $-1\leq q \leq 2$ at fixed $\lambda=0.1$. 
In Eqs. (\ref{eq:fss_tauWLl})--(\ref{eq:fss_deltaWLl}) only two scaling functions are present, $\mathcal{T}_q$ and 
$\mathcal{A}_q$, therefore we investigated $\tilde{D}_q(W,L,\lambda=0.1)$ and $\tilde{\alpha}_q(W,L,\lambda=0.1)$  only
using ensemble and typical averaging (see Sec.~\ref{sec:MFSS}).

In order to fit the scaling law (\ref{eq:fss_anderson_scalinglaw}) we used MINUIT. To find the best fit to the data obtained numerically the 
order of expansion of $\mathcal{G}^{rel/irrel}_q$, $\varrho$ and $\eta$ must be decided by choosing the values of $n_{rel}, n_{irrel}, n_{\varrho}$ 
and $n_{\eta}$. Since the relevant operator is more important than the irrelevant one we always used $n_{rel}\geq n_{irrel}$ and $n_{\varrho}\geq n_{\eta}$. 
To choose the order of the expansion we used basically three criteria. The first criterion we took into account was how close the ratio $\chi^2/({\rm NDF}-1)$ 
approached one. $\chi^2$ is the sum of the squared differences between the data points and the best fit weighted by the inverse variance of the 
data points, and $NDF$ is the number of degrees of freedom, namely the number of data points minus the number of fit parameters. 
A ratio $\chi^2/({\rm NDF}-1)\approx 1$ means, that the deviations from the best fit are in the order of the standard deviation. The second criterion was, 
that the fit has to be stable against changing the expansion orders, i.e. adding a few new expansion terms. From the fits that fulfilled the first two criteria 
we chose the simplest model, with the lowest expansion orders. Sometimes we also took into account the error bars, and we chose the model with the 
lowest error bar for the most important quantities ($W_c,\nu$, {\it etc}...), if similar models fulfilled the first two criteria.

The error bars of the best fit parameters were obtained by a Monte-Carlo simulation. The data points are results of averaging, so due to central limit 
theorem they have a Gaussian distribution. Therefore we generated Gaussian random numbers with parameters corresponding the mean and 
standard deviation of the raw data points and then found the best fit. Repeating this procedure $N_{\rm MC}=100$ times provided us the distribution 
of the fit parameters. We chose $95\%$ confidence level to obtain the error bars. We performed FSS for 
$\tilde{D}_q^{ens},\tilde{D}_q^{typ},\tilde{\alpha}_q^{ens}$ and $\tilde{\alpha}_q^{typ}$.

The results were very similar to the ones obtained by Rodriguez {\it et al.}~\cite{Rodriguez11}. In the  $q$-range we investigated the results 
were $q$-independent for $\tilde{D}_q^{ens},\tilde{D}_q^{typ},\tilde{\alpha}_q^{ens}$ and $\tilde{\alpha}_q^{typ}$ within $95\%$ confidence interval. 
The numerical values of $W_c$, $\nu$ and $y$ have been obtained in excellent agreement with the results of Ref.~\onlinecite{Rodriguez11}. 
Hence we concluded, that our method has been confirmed. The disadvantage of this method is, that the constant term of $\mathcal{G}_q$ does not equal 
to the corresponding MFE, since $\lambda$ is fixed instead of tending to zero. It would be possible to perform multifractal finite size scaling 
(MFSS) at different $\lambda$-s, and then obtain the MFEs for $\lambda \to 0$.

\subsubsection{Finite size scaling at fixed $\ell$}
\label{sec:fss_anderson_l}
The main goal of the present work is to investigate the quantum percolation problem, where a fraction of lattice points is missing. In this case performing 
the coarse graining technique defined above immediate difficulties arise. It is not clear how the $\ell$-sized boxes have to be made, or how the boxes 
containing different number of filled sites should be compared. One way to resolve this problem is to choose $\ell=1$, meaning that a box contains 
only one site. Eventhough this choice eventually opens the possibility to extend the MFSS method for irregular lattices or even for graphs and networks
in the future, there is also a huge cost to be paid: the smoothing effect of the coarse graining is lost, and only the more complicated method of fixed-$\ell$
technique described in Sec.~\ref{fss_fixed_l} remains.

There is always some numerical noise on the data, which becomes even more relevant for the smallest wave function components. In case of negative $q$ 
these uncertain small values are dominating the sums in $R_q$ and $S_q$ (see. Eqs.~(\ref{eq:SqRq})). Coarse graining clearly suppresses this effect, 
because for $\ell>1$ in an $\ell\times\ell\times\ell$ sized box positive and negative errors can cancel each other. Another effect is, that in a box large and small wave function amplitudes appear together with high probability. This way the relative 
error of a $\mu_k$ box probability is reduced with coarse graining, in other words coarse graining has a nice smoothing effect. At fixed $\ell=1$ 
this effect is missing, thus for $q<0$ the numerically obtained $\tilde{D}_q^{ens},\tilde{D}_q^{typ},\tilde{\alpha}_q^{ens}$ and $\tilde{\alpha}_q^{typ}$ 
(see e.g. Eqs.~(\ref{eq:fss_FSGMFE_sub1})--(\ref{eq:fss_FSGMFE_sub4})) values are very noisy. 
This makes every attempt to get results for negative $q$ very hard if not impossible. 

The other problem is, that the scaling law becomes more complicated, the leading number of fit parameters are growing as $\sim n_{rel/irrel}^2$ for 
fixed $\ell=1$, instead of $\sim n_{rel/irrel}$ as for fixed $\lambda$.

Performing the MFSS another problem appeared with Eq.~(\ref{eq:fss_anderson_Gammal}). During the fit the irrelevant exponent, $y$, converged 
to very small $(10^{-3}-10^{-5})$ or very large $(10^2-10^3)$ values. In the first case the irrelevant term can be merged with the relevant one, since $\eta$ is in 
most cases constant. In the second case $L^{-y}$ suppresses the irrelevant part. This caused really large errors in the $b_{ij}$, and made the whole 
irrelevant part meaningless.

To find out whether this is just a numerical problem or there is also some systematic physical reason behind this behavior we modeled the above problem: 
First a dataset was made by evaluating the expression (\ref{eq:fss_anderson_Gammal}) at system sizes and disorder we used before, with some expansion 
parameter values similar to the ones provided by previous MFSS procedures. Of course fitting Eq.~(\ref{eq:fss_anderson_Gammal}) to this dataset gave a 
perfect fit. Now adding some small random noise to the initial dataset started to shift the resulting fit parameters a little. By increasing the noise to the order 
of the standard deviation of the original dataset for the Anderson model the fit showed the expected phenomenon: The irrelevant exponent, $y$, converged 
to either large or small values. This shows, that this is just a numerical artifact. There is a shift on the $\tilde{D}_q(W,L)$ curves for different system 
sizes, see Fig.~\ref{fig:qperc_fss_alphaqDq_raw}. 
This shift comes mainly from the $1/\ln L$ term in Eq.~(\ref{eq:fss_anderson_Gammal}), and if noise is present it is numerically hard to determine the effect 
of the $L^{-y}$ irrelevant part. All in all, however, in a finite system irrelevant operators are always present, considering an irrelevant term will only increase 
the error of the fit parameters. Therefore it seems to be useful to drop the irrelevant part, and keep the relevant one only. This way the fitting function reads as
\begin{equation}
\tilde{G}_q(W,L)=G_q+\frac{1}{\ln L}\left( \sum_{i=0}^{n_{rel}}\sum_{j=0}^{i} a_{ij}\varrho^i L^{\frac{j}{\nu}} \right).
\label{eq:fss_fitformulra_l_noy}
\end{equation}

We performed MFSS in the range $0\leq q \leq 2$ with this formula at fixed $\ell=1$ for the Anderson model. Similarly to the case of fixed $\lambda$ at fixed energy, 
$E$, and $q$ one has to decide the order of the Taylor-expansion of the $\mathcal{G}$ scaling function. To do this we used similar criteria as before. 
The only difference was, that unfortunately the fits were not so stable against changing the expansion orders, $n_{rel}$ and $n_{\rho}$, as the ones for 
fixed $\lambda$, because at fixed $\ell=1$ we had to fit much more parameters to the same amount of data. The value of the critical point must be 
$q$-independent, which -- contrary to the case of fixed $\lambda$ -- we had to keep also as a criterion. We had to compare fits at different $q$ values 
and choose the lowest expansion orders that led to a $q$-independent critical point, and still had $\chi^2/({\rm NDF}-1)$ ratio close to one. 
In some cases we also had to leave out the smallest system size(s), i.e. choose $L_{min}=30$ or $40$ instead of $20$ to fulfill the criteria above.  

The results were acceptable only 
approximately in the range $0\leq q \leq 1$. If $q\geq 1$ fit parameters started to shift, sometimes out of the confidence band of those obtained 
for smaller $q$ values, and error bars were growing extremely large. Similar effects of growing error bars for $q\geq 1$  has been seen earlier, 
on a moderate level at fixed $\lambda=0.1$, where the help of the smoothing effect of coarse graining is present. The reason behind 
this is, that increasing $q$ increases the numerical and statistical errors through the $\mu_k^q$ expression. As mentioned above, increasing 
error on the data makes it really difficult to get acceptable results from the MFSS.

As a result, in the range $0\leq q \leq 1$ the critical point, $W_c$, and the critical exponent, $\nu$, were found to be consistent with our results 
at fixed $\lambda=0.1$ and based on the $D_q$ and $\alpha_q$ exponents also with high precision result of Rodriguez {\it et. al}~\cite{Rodriguez11}. 
We observe the expected symmetry (\ref{eq:multifractals_Deltaalphasymmety}) for $\Delta_q$ and $\alpha_q$, our resulting MFEs fulfill these 
conditions in the range $0\leq q \leq 1$.

Summarizing the results it is possible to perform an MFSS at fixed $\ell=1$ and achieve good agreement with previous high precision 
results~\cite{Rodriguez11}. There are certainly numerical difficulties, however, that lead us to resort to the limited range of $0\leq q \leq 1$ only, 
but with further averaging the widening of this $q$-range seems to be possible.

\subsection{Numerical calculations for the 3D quantum percolation model using GMFEs}
\label{sec:fss_qperc}
The main goal in the present study was to find the mobility edge and the critical exponent of the 3D quantum percolation model, and investigate the 
multifractal properties of the critical wave functions. Since the Hamiltonian Eq.~(\ref{eq:qperc_qperchami}) is symmetric for $E\leftrightarrow -E$ exchange, 
the $E\geq 0$ interval is investigated only. We used the same numerics as in Sec~\ref{sec:fss_anderson_l}. To avoid the frequent molecular states 
(see Sec.~\ref{sec:qperc_dos}), and cover the most interesting regions of the band we chose the following energies: 
$E=0.001,\ 0.01,\ 0.1, \ 0.3,\ 0.7,\ 1.1,\ 1.5,\ 2.1,\ 3.1,\ 4.1$. For averaging we considered only one wave function per realization with the eigenvalue closest 
to the chosen energy $E$ to avoid correlations. We only used an eigenfunction if its energy was in a $\Delta E=0.01$ wide interval around $E$, except for 
$E=0.001$ and $E=0.01$, where $\Delta E=0.00001$ and $\Delta E=0.001$ were used. 

Our $\Delta E$ energy intevals are so small, that it completely excludes the effect of molcecular states. We ran a test after the finite size scaling was performed: Molecular states have strict energy value, therefore at fixed system size, $L$, disorder, $p$, and energy, $E$, we left out from our raw dataset all the wave functions  
with the same energy value (at most 2\% of the original raw dataset). Note, that these states are not necessarily molecular states, they can be regular ones, too, 
having the same energy within numerical precision. We redid our whole finite size scaling procedure (as described below), but this additional refinement had
no effect on the results. This test ensures, that we filtered out the molecular states very effectively, and if they were present in our raw dataset, their effect 
would be negligible.

At every energy we searched for the critical point
$p_c^{\scriptscriptstyle Q}$. From the approximately $\Delta p=0.01$ wide neighborhood around $p_c^{\scriptscriptstyle Q}$ we picked about $20$ values of $p$.
For higher $p_c^{\scriptscriptstyle Q}$ values at fixed system size, $L$, there are more sites in the giant cluster, thus the Hamiltonian matrix is larger, 
and it takes more time to find the closest eigenvalue to the given energy. On the other hand $R_q$ and $S_q$ are calculated from more data, 
thus they are more precise. Considering these arguments we investigated system sizes and number of samples listed in Tab.~\ref{tab:qperc_fss_systemsize}.
Altogether $45,045,000$ wave functions were calculated.

\begin{table}[h]
	\begin {center}
	\begin{tabular}{|c| c| c|}
	\hline
	\multirow{2}{*}{system size $(L)$} & \multicolumn{2}{c|}{number of samples}\\ \cline{2-3}
	 & $p_c^{\scriptscriptstyle Q}<0.41$ & $p_c^{\scriptscriptstyle Q}>0.41$ \\ \hline
	20 & 50000 & 50000 \\ \hline
	30 & 50000 & 50000 \\ \hline
	40 & 50000 & 50000 \\ \hline
	60 & 50000 & 25000\\ \hline
	80 & 20000 & 10000\\ \hline
	100 & 10000 & 5000\\ \hline
	120 & 5000 & \\ \hline
	140 & 4000 & \\ \hline
	\end{tabular}
	\caption{System sizes and number of samples of the simulation for the 3D quantum percolation model.}
	\label{tab:qperc_fss_systemsize}
	\end{center}
\end{table}

\begin{figure*}
	\begin {center}
	\begin{tabular}{c c}
	\begin{overpic}[type=pdf,ext=.pdf,read=.pdf,width=.48\linewidth]{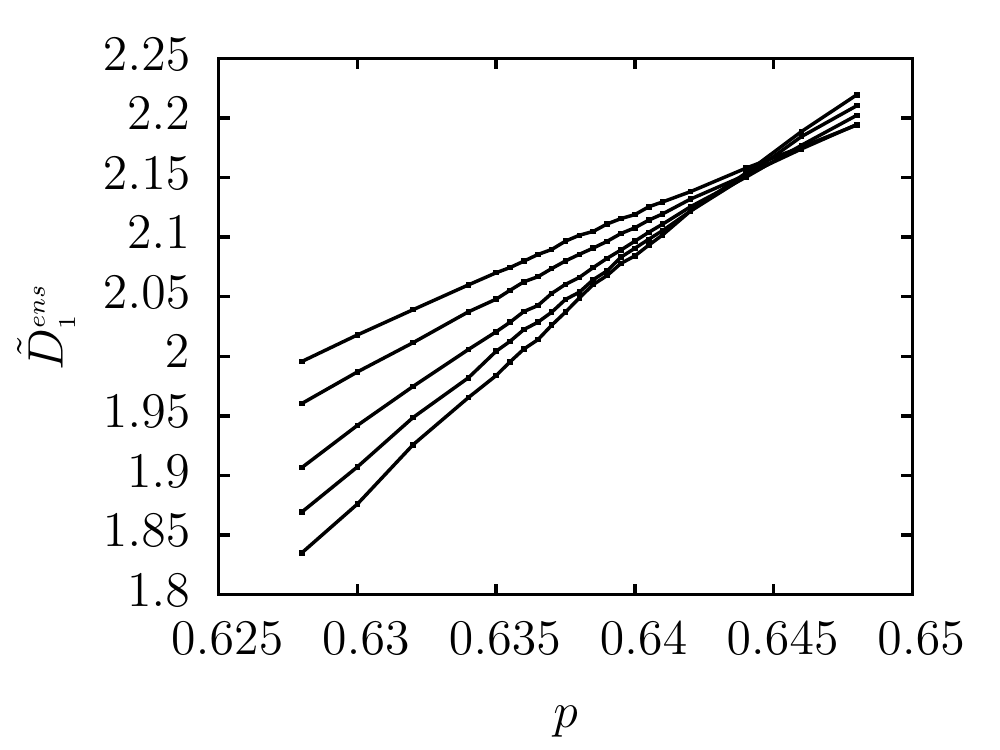}
	\put(0,70){(a)} 
	\end{overpic} & 
	\begin{overpic}[type=pdf,ext=.pdf,read=.pdf,width=.48\linewidth]{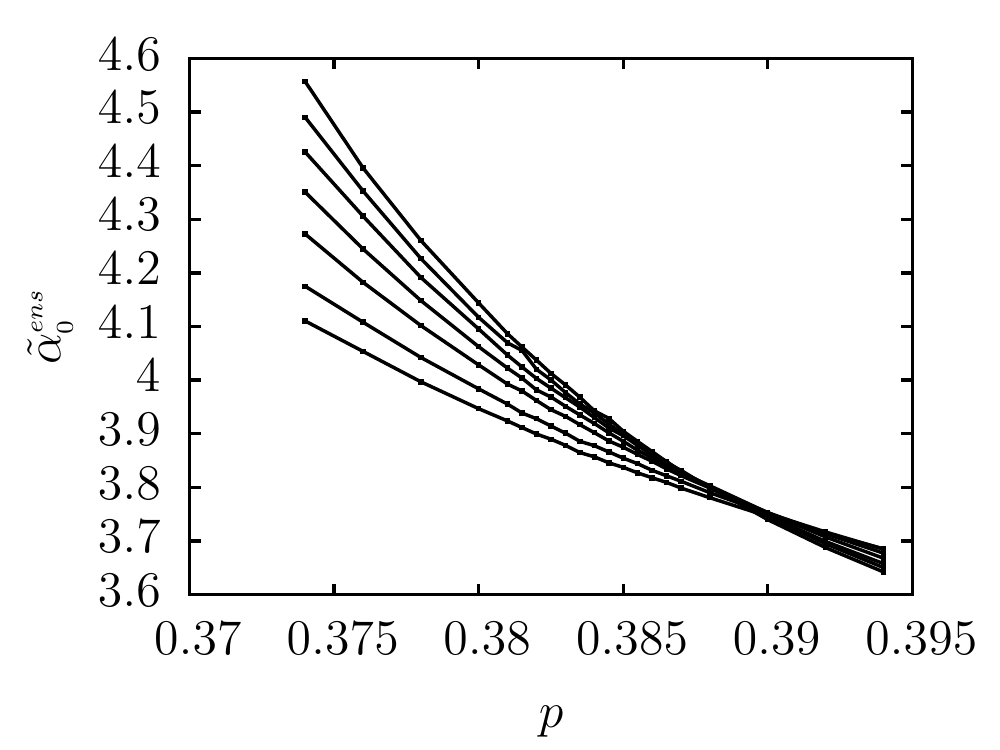} 
	\put(0,70){(b)} 
	\end{overpic} \\
	\end{tabular}
	\caption{The generalized multifractal exponents (a) $\tilde{\alpha}_{1}^{ens}(p,L,\ell=1)$ at $E=0.7$ and (b) $\tilde{D}_{0.5}^{ens}(p,L,\ell=1)$ at 
	$E=0.1$ for the 3D quantum percolation model. Points with error bars are the raw data, red solid lines are the best fits of the function 
	Eq.~\ref{eq:fss_fitformulra_l_noy} as a function of disorder, $p$, at different system sizes, $L$.}
	\label{fig:qperc_fss_alphaqDq_raw}	
	\end{center}
\end{figure*}

The method we used here has been described in Sec~\ref{sec:fss_anderson_l}. We experienced, that for typical averaging finite size scaling sometimes 
showed difficulties to converge, therefore we used the ensemble averaged exponents, $D_q^{ens}$ and $\alpha_q^{ens}$ only. The typical behavior 
of these exponents is presented in Fig.~\ref{fig:qperc_fss_alphaqDq_raw}, note that curves do not have a common crossing point due to the $1/\ln L$ term 
in Eq.~(\ref{eq:fss_fitformulra_l_noy}).

\begin{figure*}
	\begin {center}
	\begin{tabular}{c c}
	\begin{overpic}[type=pdf,ext=.pdf,read=.pdf,width=.5\linewidth]{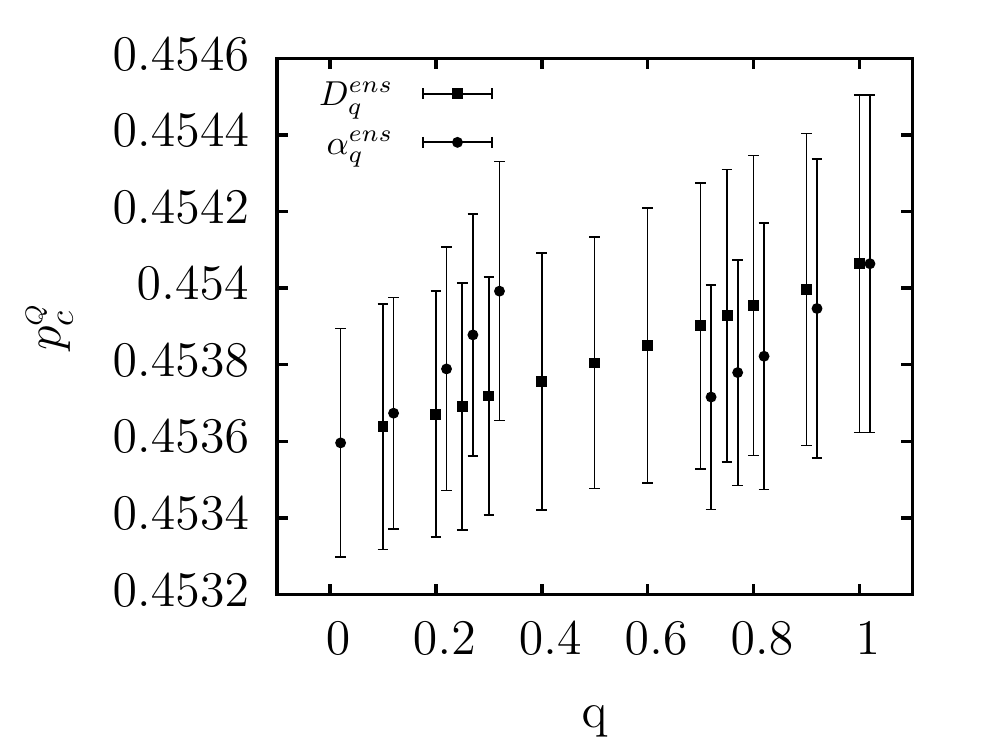} 
	\put(0,70){(a)} \end{overpic} & 
	\begin{overpic}[type=pdf,ext=.pdf,read=.pdf,width=.5\linewidth]{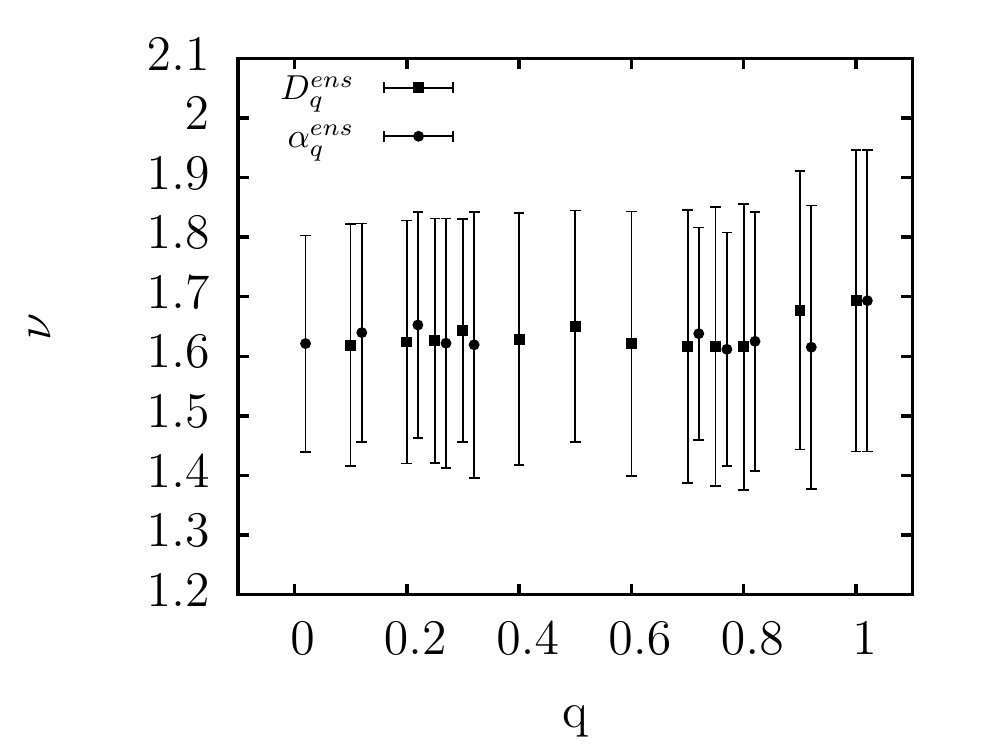} 
	\put(0,70){(b)} \end{overpic}\\
	\begin{overpic}[type=pdf,ext=.pdf,read=.pdf,width=.5\linewidth]{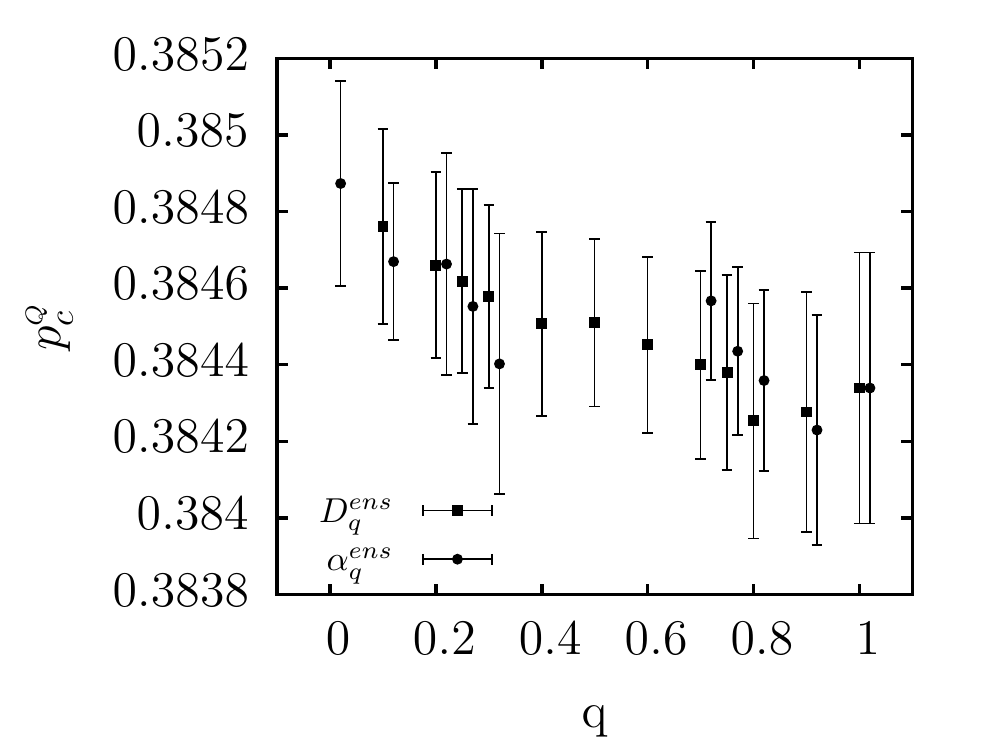} 
	\put(0,70){(c)} \end{overpic}& 
	\begin{overpic}[type=pdf,ext=.pdf,read=.pdf,width=.5\linewidth]{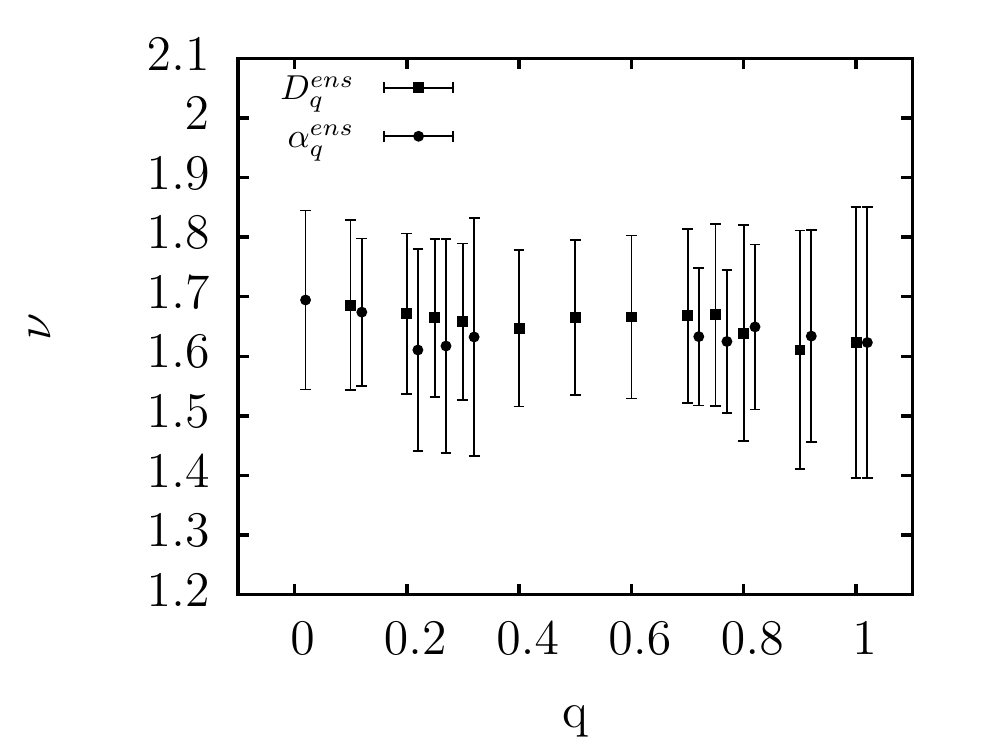} 
	\put(0,70){(d)} \end{overpic}\\
	\begin{overpic}[type=pdf,ext=.pdf,read=.pdf,width=.5\linewidth]{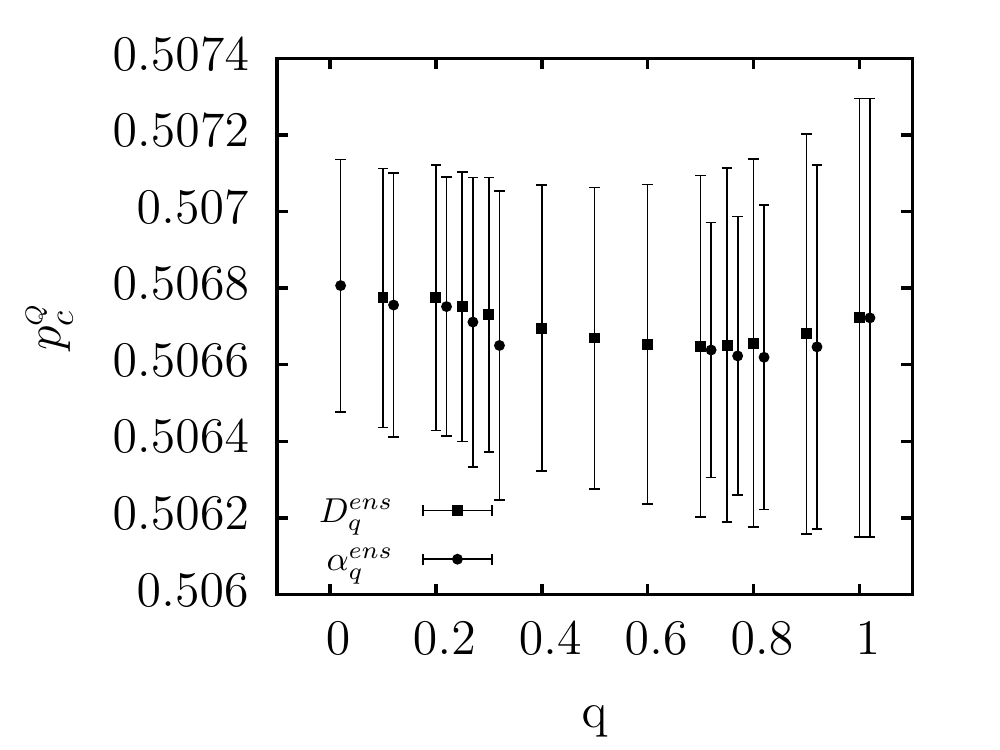}
 	\put(0,70){(e)} \end{overpic}& 
	\begin{overpic}[type=pdf,ext=.pdf,read=.pdf,width=.5\linewidth]{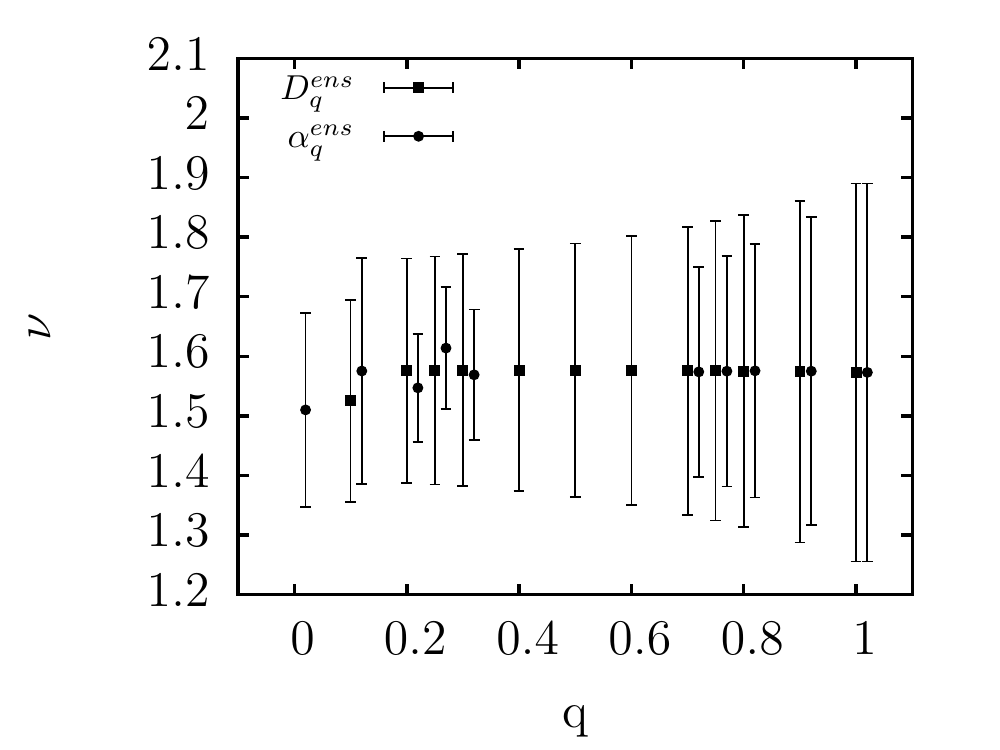}
	\put(0,70){(f)} \end{overpic} \\
	\end{tabular}
	\caption{Critical point (left column) and critical exponent (right column) of the 3D quantum percolation model at (a) and (b) $E=0.1$, 
	(c) and (d) $E=0.7$, (e) and (f) $E=3.1$. Error bars represent $95\%$ confidence levels.}
	\label{fig:qperc_pq_nu_E}	
	\end{center}
\end{figure*}

The MFSS at fixed $\ell=1$ for the range $0\leq q \leq 1$ provided critical points, critical exponents and MFEs for every $q$ value at every chosen energy, 
$E$. For fixed energy the critical points and critical exponents should be $q$-independent, which can be fulfilled within the $95\%$ confidence level, see 
Fig.~\ref{fig:qperc_pq_nu_E}.

The critical point, $p_c^{\scriptscriptstyle Q}$, shifts in most cases, but the shift is within the $95\%$ confidence band. An interesting feature is, that
$p_c^{\scriptscriptstyle Q}$ obtained from $\alpha_q$ for $q\leq 0.5$ and $q\geq 0.5$ shifts in the opposite direction. For $\alpha_{0.4}$ and $\alpha_{0.6}$ 
the MFSS mostly did not converge since $\alpha_{0.5}=d$ and close to the $q=0.5$ point $\tilde{\alpha}$ curves have similar steepness close to the critical 
point, therefore it is numerically hard to determine a well--defined crossing point after scaling out the $\ln L$ shift. Therefore these data are not presented 
in Fig.~\ref{fig:qperc_pq_nu_E}.

\begin{figure*}
	\begin {center}
	\begin{tabular}{c c}
	\begin{overpic}[type=pdf,ext=.pdf,read=.pdf,width=0.5\linewidth]{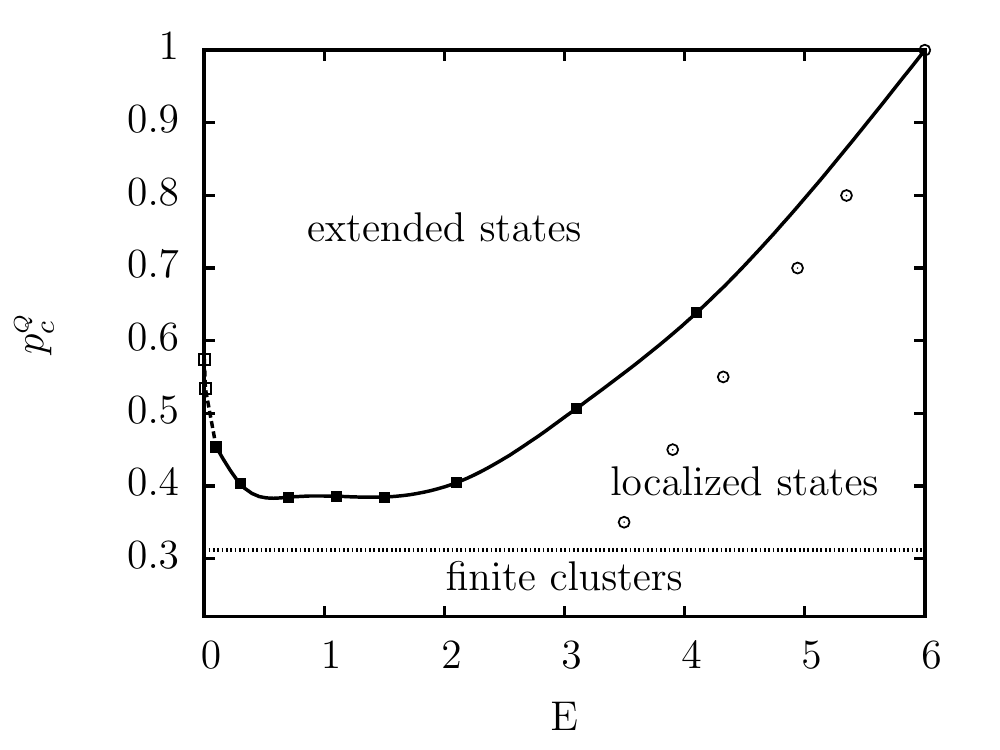} \put(0,70){(a)} \end{overpic} & 
	\begin{overpic}[type=pdf,ext=.pdf,read=.pdf,width=0.5\linewidth]{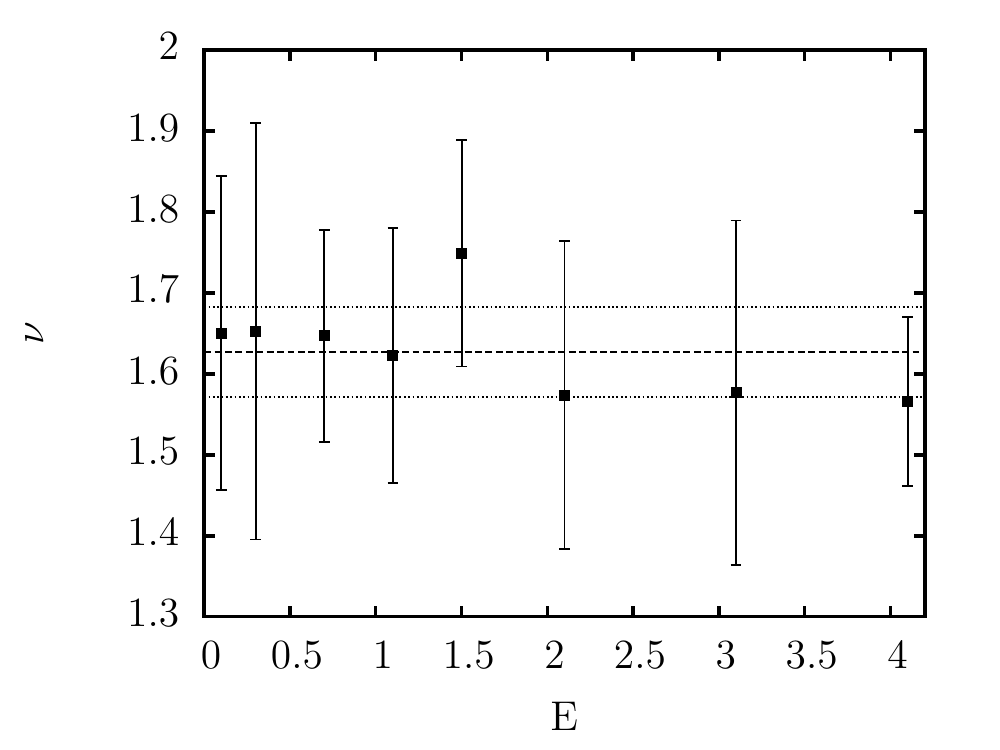} \put(0,70){(b)} \end{overpic} \\
	\end{tabular}
	\caption{(a) Mobility edge for the 3D quantum percolation model, dotted line denotes the classical percolation threshold, 
	$p_c^{\scriptscriptstyle C}=0.3116\pm0.0002$~\cite{Stauffert}. 
	Circles are approximate values of the bandwidth, beyond them only the Lifshitz-tail is present.
	Squares are results from MFSS, line is a spline to guide for the eye. 
	Empty squares and dashed lines are for approximate data obtained from MFSS at $E=0.001$ and $E=0.01$ (b) Critical exponent for the 3D quantum
	percolation model. Error bars are for $95\%$ confidence band. Dashed line is the average, dotted lines note the $95\%$ confidence band around the 
	average. The resulting critical exponent is $\nu=1.622\ (1.587..1.658)$}
	\label{fig:qperc_fss_mobedge_nu}	
	\end{center}
\end{figure*}

For $E=0.001$ and $E=0.01$ the MFSS showed severe convergence troubles, and even if it converged, provided fit parameters with very large error. 
The reason behind this behavior is presumably the close vicinity of the pseudogap at $E=0$ in the DOS, and it is very hard even to find eigenvalues close 
enough to the desired energies $E=0.001$ or $E=0.01$. Another difficulty in this case is that the mobility edge becomes anomalous approaching $E=0$, 
see Fig.~\ref{fig:qperc_fss_mobedge_nu}(a). Therefore only a narrow energy-band is permitted for averaging around $E=0.001$ or $E=0.01$, 
which decreases further the possible number of eigenstates. For these reasons parameters coming from MFSS at $E=0.001$ and $E=0.01$ were only 
used to plot the mobility edge, these two points are denoted with empty squares in Fig.~\ref{fig:qperc_fss_mobedge_nu}(a).

At fixed energy we picked one $q$ point, that represents well the results for that energy, see Tab.~\ref{tab:qperc_mobedge_data}. The 
$p_c^{\scriptscriptstyle Q}$ values are leading to a mobility edge, see Fig.~\ref{fig:qperc_fss_mobedge_nu}(a). The values of $\nu$ are independent, 
and should not depend on the energy. Thus they can be averaged, providing a more precise critical exponent $\nu=1.622\ (1.587..1.658)$, 
see Fig.~\ref{fig:qperc_fss_mobedge_nu}(b). To derive the average, the data points were weighted by their inverse variance, the error bar is twice the 
standard deviation of the mean, which is about the $95\%$ confidence band for a Gaussian.

\begin{table*}
\begin{tabular}{|c|c|c|c|c|c|c|c|c|}
\hline
$\mathbf E$ & {\bf MFE} & $\mathbf p_{\mathbf c}^{\scriptscriptstyle \mathbf Q}$ & $\boldsymbol \nu$ & $\mathbf{NDF}$ & $\boldsymbol \chi^{\mathbf 2}$ & $\mathbf L_{\mathbf{min}}$ 
& $\mathbf n_{\mathbf{rel}}$ & $\mathbf n_{\boldsymbol \rho}$\\ \hline
$0.1$ & $D_{0.5}=2.421\ (2.416..2.426)$ & $0.45384\ (0.45365..0.45402)$ & $1.591\ 1.508..1.682$ & $136$ & $113$ & $20$ & $3$ & $1$\\ \hline
$0.3$ & $D_{0.5}=2.397\ (2.393..2.402)$ & $0.40241\ (0.40228..0.40257)$ & $1.705\ 1.578..1.879$ & $157$ & $123$ & $20$ & $3$ & $1$\\ \hline
$0.7$ & $D_{0.6}=2.271\ (2.265..2.278)$ & $0.38402\ (0.38387..0.38418)$ & $1.645\ 1.572..1.741$ & $181$ & $150$ & $20$ & $4$ & $1$\\ \hline
$1.1$ & $D_{0.6}=2.262\ (2.257..2.268)$ & $0.38518\ (0.38504..0.38531)$ & $1.609\ 1.542..1.688$ & $243$ & $155$ & $20$ & $3$ & $1$\\ \hline
$1.5$ & $D_{0.8}=2.027\ (2.020..2.035)$ & $0.38459\ (0.38443..0.38476)$ & $1.688\ 1.589..1.789$ & $144$ & $154$ & $20$ & $3$ & $2$\\ \hline
$2.1$ & $D_{0.5}=2.439\ (2.431..2.448)$ & $0.40466\ (0.40443..0.40492)$ & $1.606\ 1.530..1.692$ & $127$ & $116$ & $40$ & $2$ & $2$\\ \hline
$3.1$ & $D_{0.4}=2.542\ (2.538..2.546)$ & $0.50628\ (0.50606..0.50647)$ & $1.603\ 1.515..1.695$ & $138$ & $113$ & $20$ & $3$ & $1$\\ \hline
$4.1$ & $\alpha_{0.9}=2.108\ (2.101..2.114)$ & $0.63827\ (0.63806..0.63845)$ & $1.584\ 1.486..1.699$ & $128$ & $113$ & $20$ & $3$ & $1$ \\ \hline

\end{tabular}
\caption{Resulting data along the mobility edge. These $q$ values were chosen to compute $\nu$ and obtain the mobility edge.}
\label{tab:qperc_mobedge_data}
\end{table*}

\begin{figure*}[!]
	\begin {center}
	\includegraphics[type=pdf,ext=.pdf,read=.pdf,width=.6\linewidth]{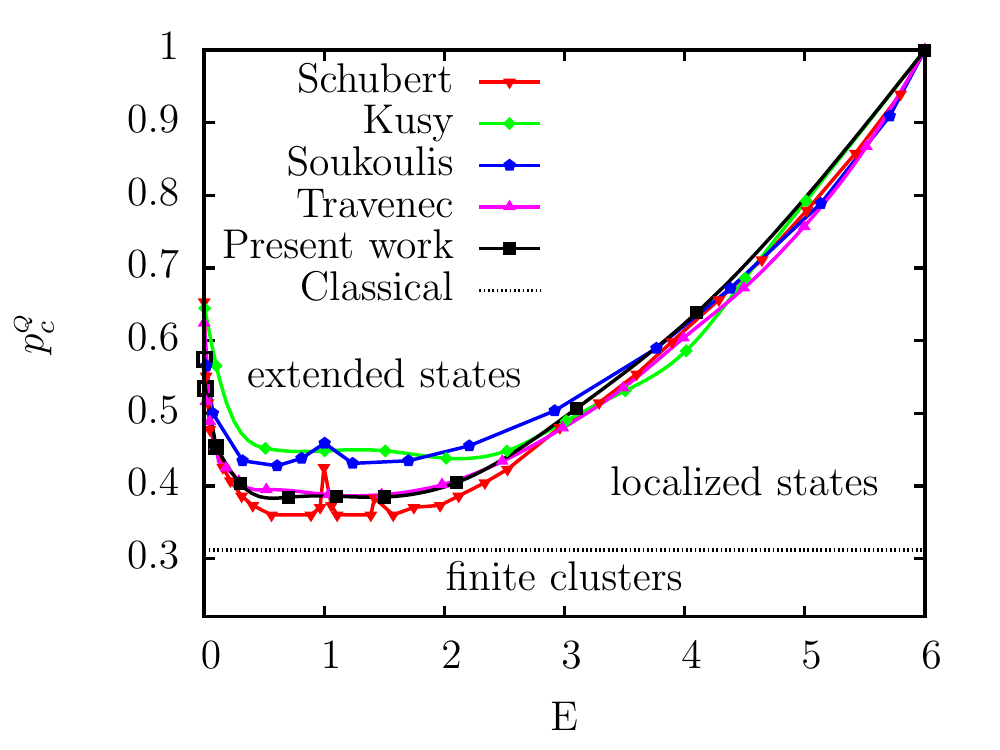}
	\caption{Mobility edge of the 3D quantum percolation model in the literature.~\cite{Schubert,Kusy,Soukoulis,Travenec,Stauffert}}
	\label{fig:qperc_fss_mobedge_literature}	
	\end{center}
\end{figure*}

In the literature there are previous works resulting mobility edge~\cite{Schubert,Kusy,Soukoulis,Travenec,Stauffert}, 
see Fig.~\ref{fig:qperc_fss_mobedge_literature}. The shape of these curves are very similar: a steep decrease around $E=0$, then a plateau resulting in
a global quantum percolation threshold for the system, and finally an increasing behavior with growing energy. The curves are in good qualitative agreement 
with each other, beyond $E=3$ quantitative agreement is also present. Curves of Soukoulis~\cite{Soukoulis} and Schubert~\cite{Schubert} have jumps at 
$E=1$ and $E=\sqrt{2}$ (only Ref.~\onlinecite{Schubert}) due to the most frequent molecular states probably. Our curve is in really good agreement with recent 
result of Travenec\cite{Travenec} obtained by transfer matrix methods, curves are almost covering each other. His critical exponent is also in good agreement 
with ours, see Tab.~\ref{tab:qperc_nu_literature}.

At low $p$ values the bandwidth is small, but increasing $p$ results in a wider band. In the Lifshitz-tail only localized states are present, 
therefore the mobility edge curve should be above the curve of the bandwidth. As a result the mobility edge curve increases at high energies in 
Fig.~\ref{fig:qperc_fss_mobedge_nu}(a). 
Reaching the edge of the band, $E\to 6$, the mobility edges drawn from the data points of different authors seem to converge to $1$. Therefore we put 
a point in the right-top corner of Fig.~\ref{fig:qperc_fss_mobedge_literature}, however, at $p=1$ the sample is a perfect crystal, and wave functions are 
completely extended Bloch-functions over the complete band.

Exactly at the center of the band, $E=0$, on the other hand, extremely localized molecular states disturb the picture, 
in addition close to the band center a pseudogap forms in the DOS (see Fig.~\ref{fig:qperc_DOS}), therefore this regime is really hard 
to investigate numerically. 
Eventhough the localized molecular states at $E=0$ belong to the point spectrum, it is still not clear, what is the $E\to 0$ limit of the mobility edge, 
describing the continuous spectrum. The question arises: Does the very steep increase of the mobility edge approaching $E=0$ result in a 
$p_c^{\scriptscriptstyle Q}(E\to0)\to 1$ or the limit is lower than one? Based on the arguments in Sec.~\ref{sec:qperc_dos} our guess is, 
that at any finite disorder, $p<1$, there are localized states near $E=0$, resulting a limit of unity for the mobility edge, $p_c^{\scriptscriptstyle Q}(E\to0)=1$.

Some values of the critical exponent can also be found in the literature. In Tab.~\ref{tab:qperc_nu_literature} we collected these values ranging from 
$1.2$ to $1.95$. Because of the more limited computational efforts, previous works used much smaller system sizes compared to our possibilities, 
leading to much bigger finite size effects, affecting their FSS. Conductivity or transfer matrix methods used to overestimate, while level statistics and 
Green-function techniques used to underestimate the critical exponent, $\nu$. Our critical exponent is practically in the center of the interval of previous 
results $1.2\leq \nu\leq 1.95$. Our exponent, $\nu=1.622\ (1.587..1.658)$ is in very good agreement with the most recent study of Travenec\cite{Travenec} 
similarly to the mobility edge. Furthermore the critical exponent is within confidence band with our previous result for the Anderson-model at fixed $\ell=1$ 
obtained from $D_{0.6}^{ens}$ ($\nu=1.617\ (1.485..1.783)$) or at fixed $\lambda=0.1$ obtained from $\alpha_{0.6}^{ens}$ ($\nu=1.598\ (1.576..1.616)$) even 
further with the high precision value ($\nu=1.590\ (1.579..1.602)$) of Rodriguez {\it et. al}~\cite{Rodriguez11}, however our result seems to be a bit higher. 
Based on these facts our work provides further evidence to previous conjectures and statements saying, that the Anderson model and the 3D quantum 
percolation model belong to the same universality class.
\vfill

\begin{table*}
\begin{tabular}{|c|c|c|c|c|}
\hline
{\bf Author} & {\bf Year} & $\mathbf{\nu}$ & {\bf Method} & {\bf Sytem size}\\ \hline
Root-Bauer-Skinner\cite{Root} & 1988 & $1.8\pm0.11$ & conductivity & $L=3-9$\\ \hline
Koslowski-von Niessen\cite{Koslowski} & 1991 & $1.95\pm0.12$ & conductivity & $L=6-9$\\ \hline
Berkovits-Avishai\cite{Berkovits} & 1996 & $1.35\pm0.1$ & level statistics & $L=7-15$\\ \hline
Kusy {\it et al.}\cite{Kusy} & 1997 & $1.2\pm0.2$ & Green-function & $L=4-8$\\ \hline
Kaneko-Ohtsuki & 1999 & $1.46\pm0.09$ & level statistics & $L=12-21$ \\ \hline
Travenec\cite{Travenec} & 2008 & $1.6\pm 0.1$ & conductivity & $L=14-20$\\ \hline
Present work & 2014 & $1.622\pm 0.035$ & multifractality & $L=20-140$\\ \hline
\end{tabular}
\caption{Critical exponent of the 3D quantum percolation model in the literature.}
\label{tab:qperc_nu_literature}
\end{table*}

\section{Analysis of MFEs of the 3D quantum percolation method}
\label{sec:gmfe_qperc}
\begin{figure*}
	\begin {center}
	\begin{tabular}{c c}
	\begin{overpic}[type=pdf,ext=.pdf,read=.pdf,width=0.5\linewidth]{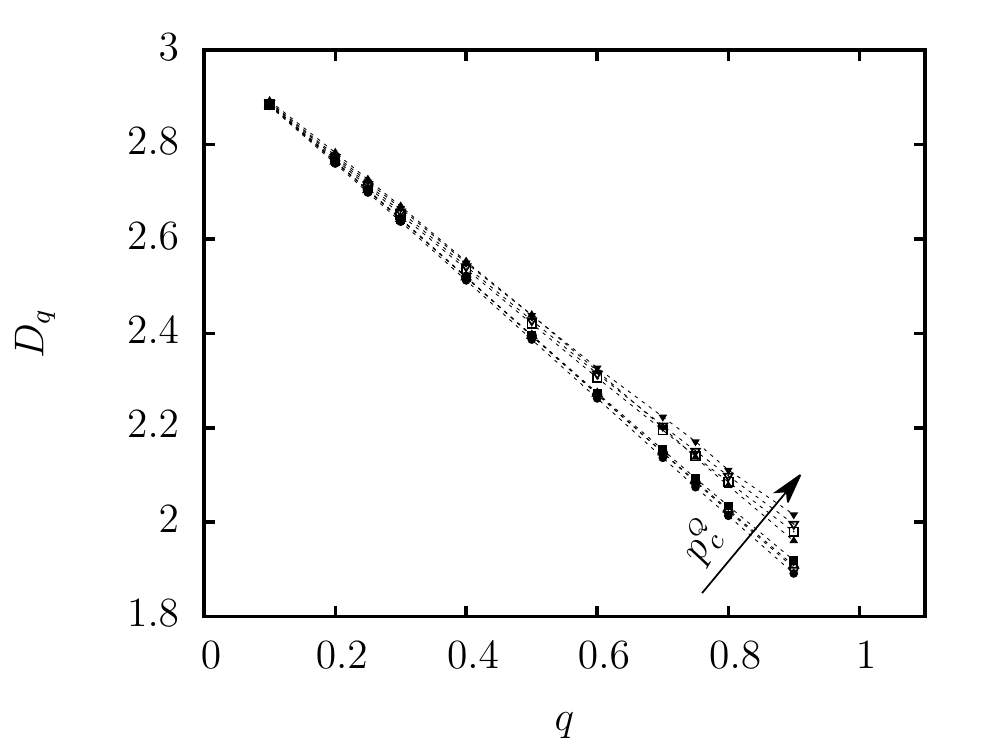}
	\put(0,70){(a)}\end{overpic} & 
	\begin{overpic}[type=pdf,ext=.pdf,read=.pdf,width=0.5\linewidth]{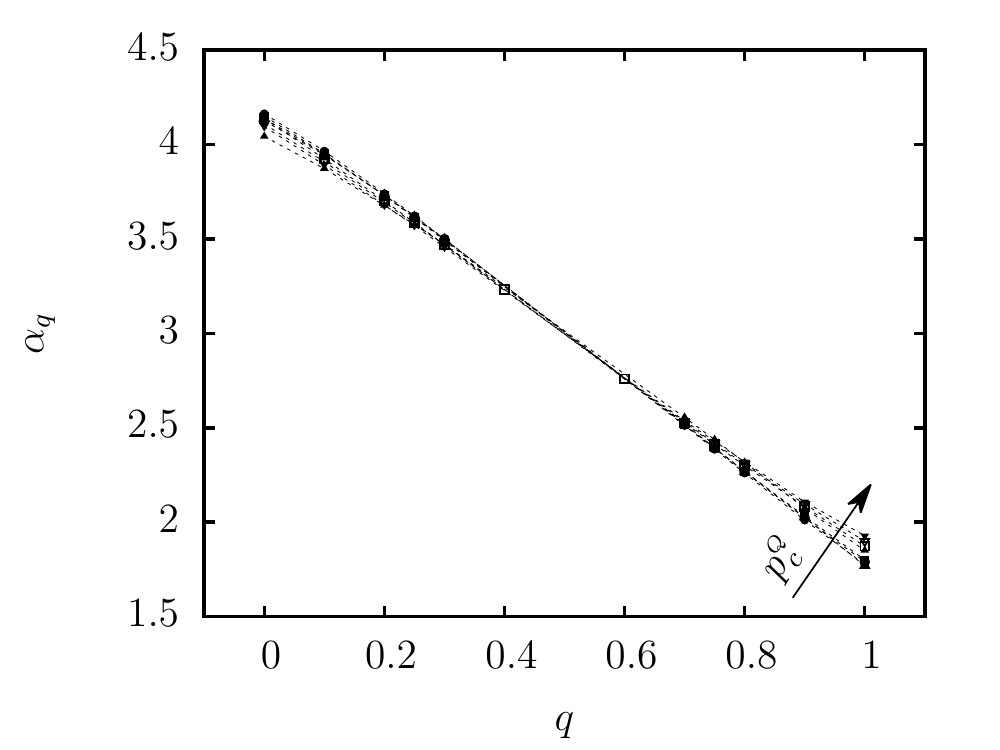} \put(0,70){(b)}\end{overpic} \\
	\begin{overpic}[type=pdf,ext=.pdf,read=.pdf,width=0.5\linewidth]{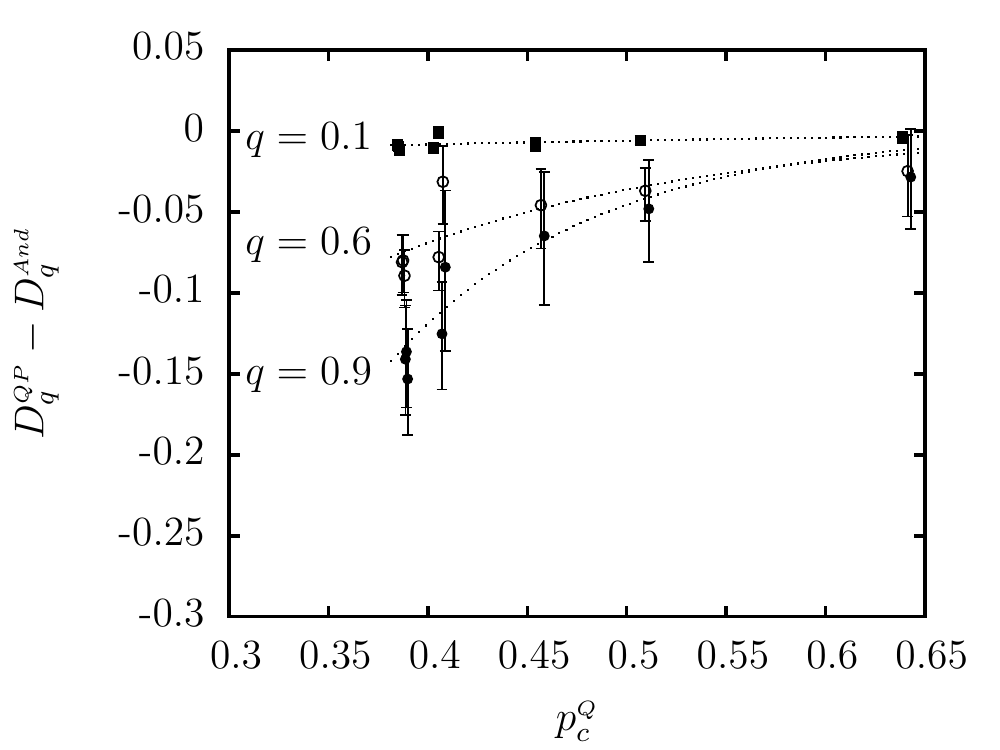}	
		\put(0,70){(c)}
		\put(52,13.8){\includegraphics[type=pdf,ext=.pdf,read=.pdf,width=0.195\linewidth]{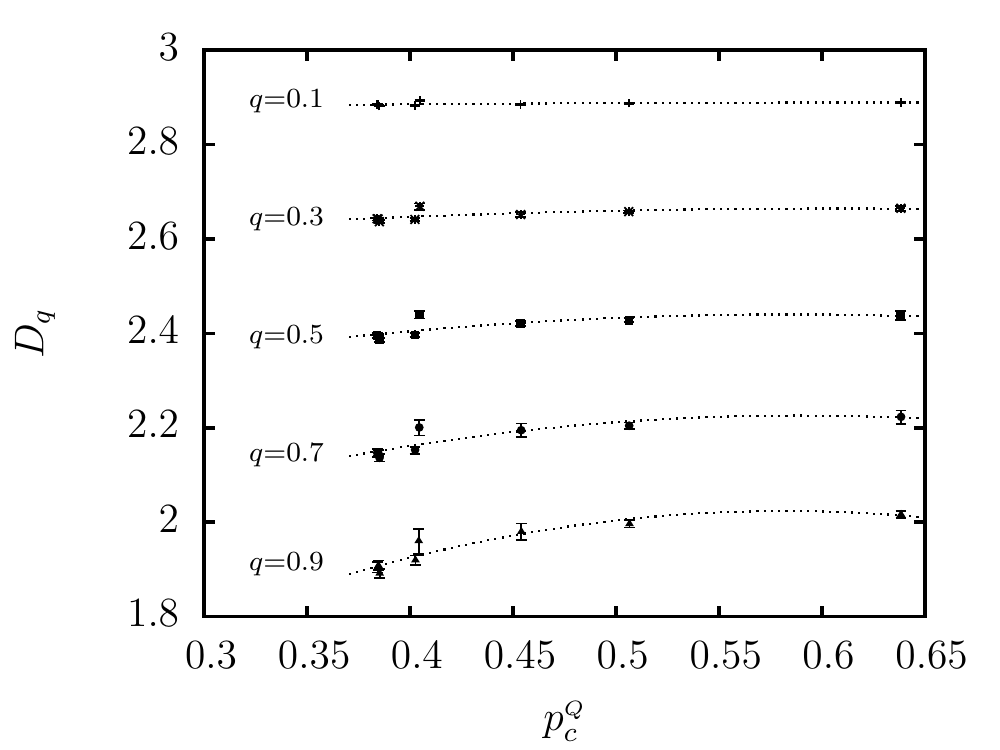}}	
		\end{overpic} & 
	\begin{overpic}[type=pdf,ext=.pdf,read=.pdf,width=0.5\linewidth]{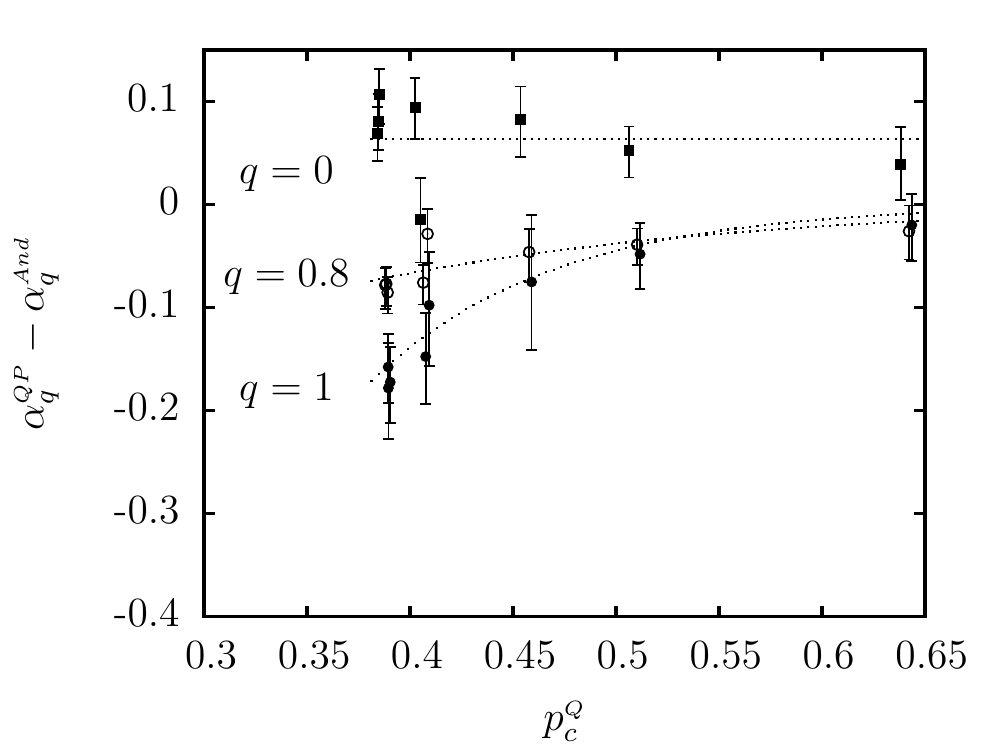} 
		\put(0,70){(d)}
		\put(52,13.8){\includegraphics[type=pdf,ext=.pdf,read=.pdf,width=0.195\linewidth]{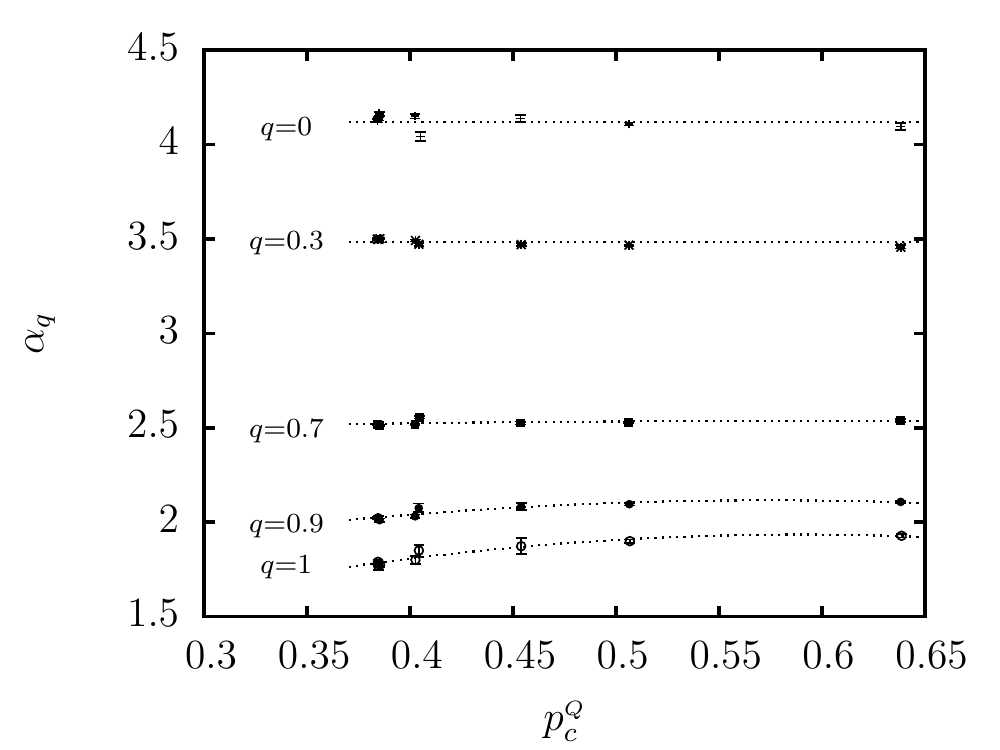}}
		\end{overpic} \\
	\end{tabular}
	\caption{First row: (a) $D_q$ (b) $\alpha_q$ as a function of $q$ at different energies, $E$. 
	Second row: GMFEs shifted by their value for the Anderson model, (a) $D_q^{QP}-D_q^{And}$ (b) $\alpha_q^{QP}-\alpha_q^{And}$ as a function of 
	$p_c^{\scriptscriptstyle Q}$ at different $q$ values. Dotted lines are guides for the eye, error bars represent a $95\%$ confidence band on (c) and (d). 
	Insets are the same, but without the shift.}
	\label{fig:qperc_Dq_alphaq_shift}
	\end{center}
\end{figure*}

\begin{figure*}
\begin {center}
	\begin{tabular}{c c}
	\begin{overpic}[type=pdf,ext=.pdf,read=.pdf,width=0.5\linewidth]{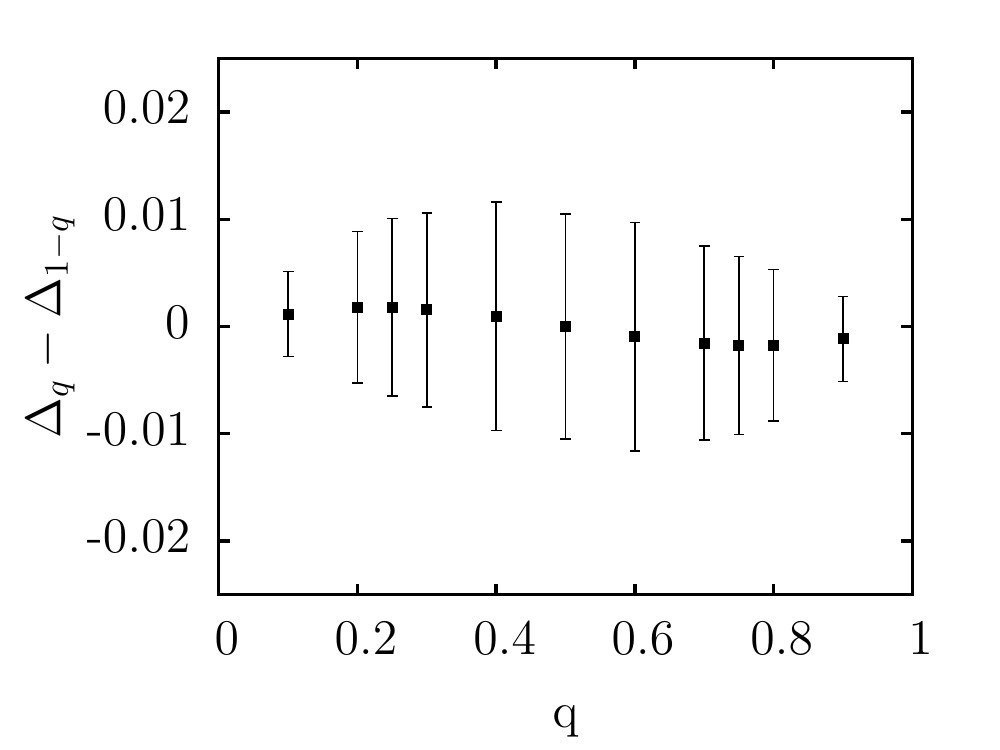} 
	\put(0,70){(a)} \end{overpic} & 
	\begin{overpic}[type=pdf,ext=.pdf,read=.pdf,width=0.5\linewidth]{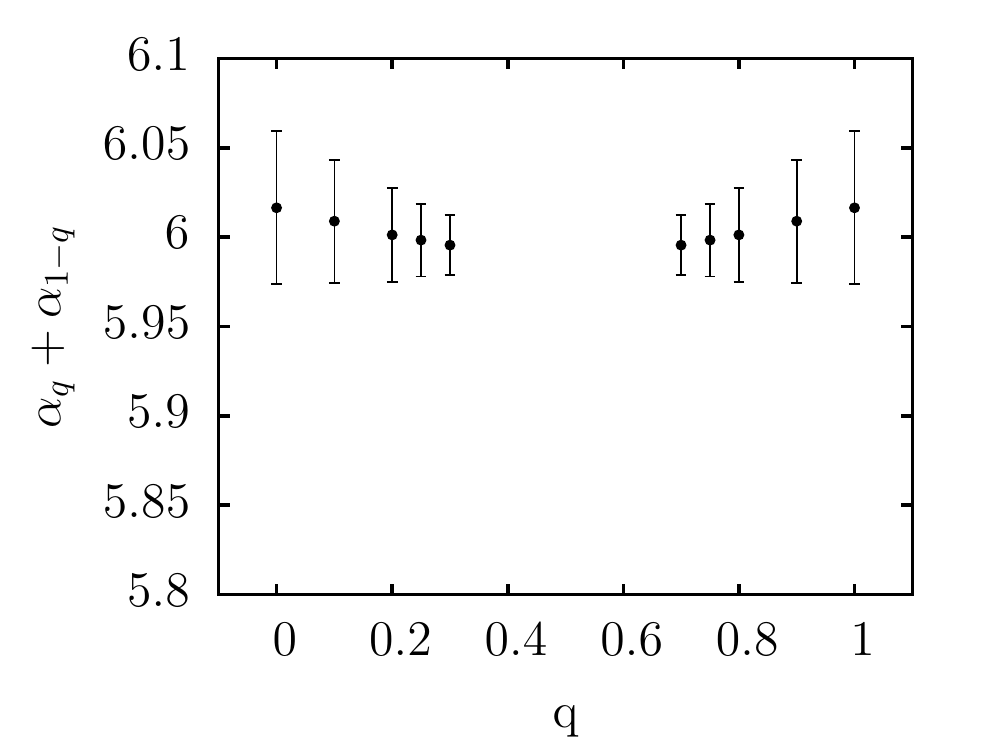} 
	\put(0,70){(b)} \end{overpic} \\
	\begin{overpic}[type=pdf,ext=.pdf,read=.pdf,width=0.5\linewidth]{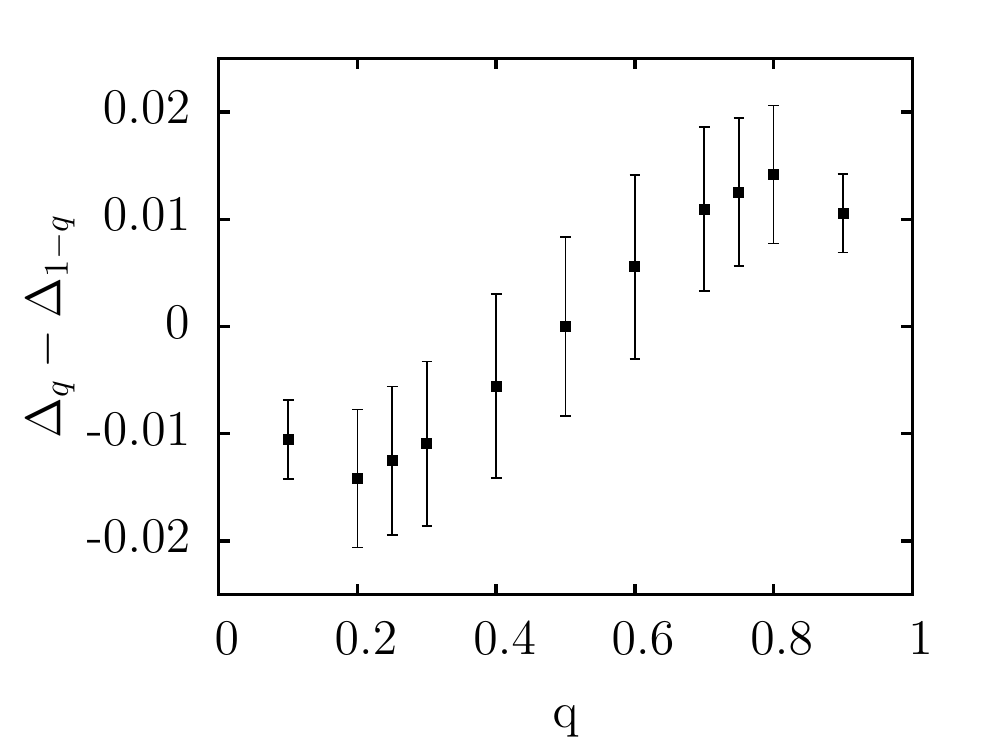}
	\put(0,70){(c)} \end{overpic} & 
	\begin{overpic}[type=pdf,ext=.pdf,read=.pdf,width=0.5\linewidth]{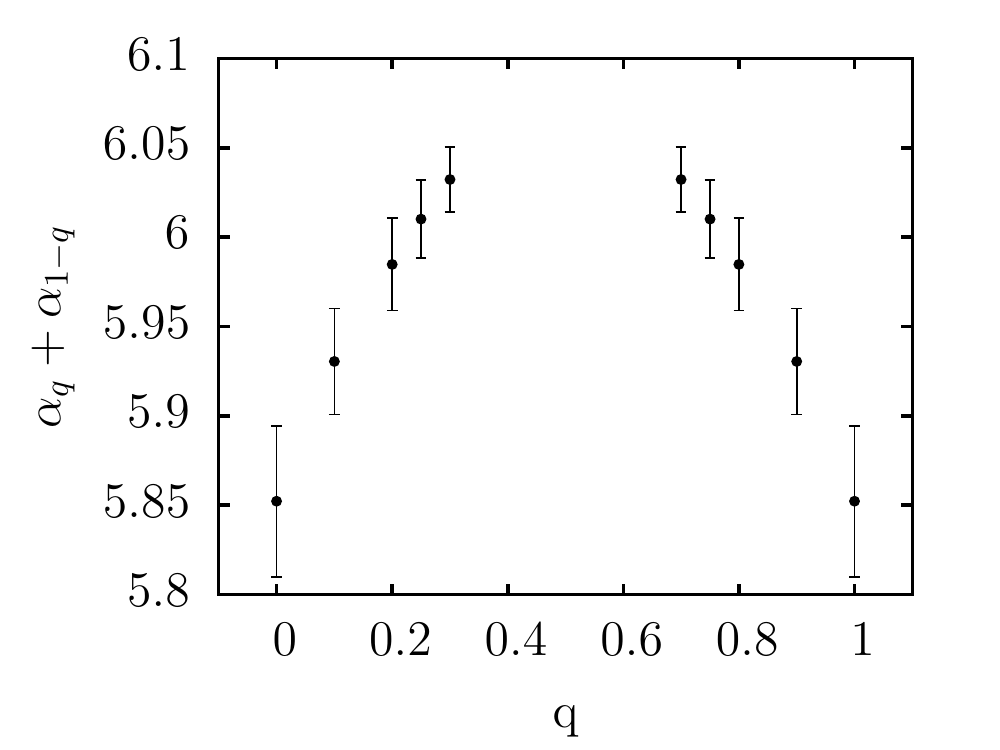}
	\put(0,70){(d)} \end{overpic} \\
	\begin{overpic}[type=pdf,ext=.pdf,read=.pdf,width=0.5\linewidth]{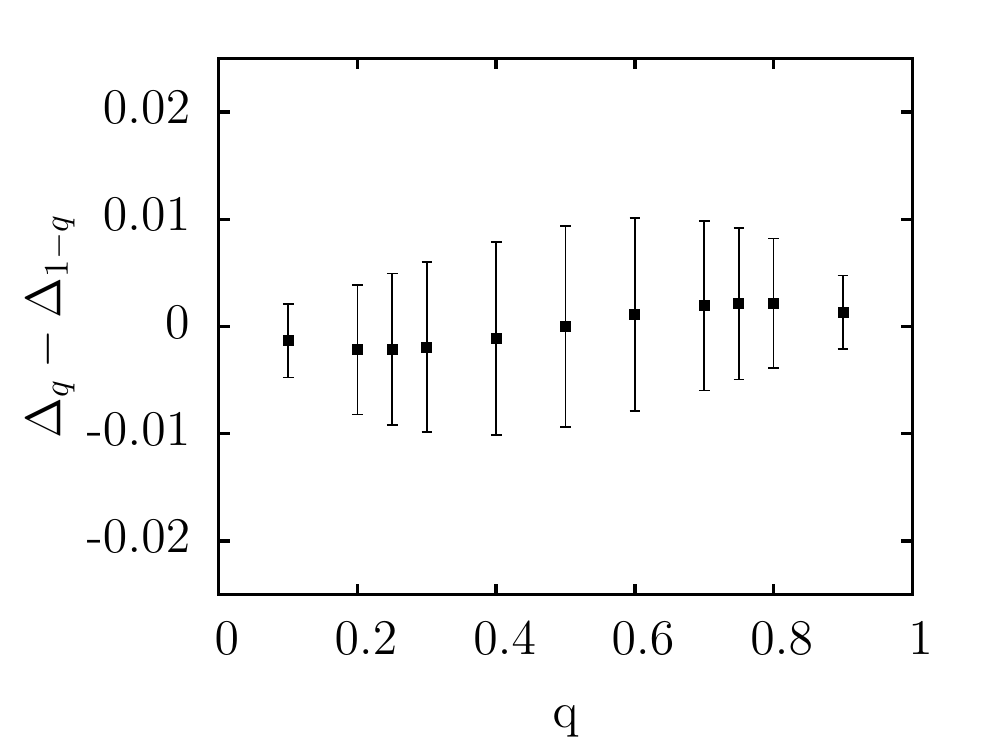}
	\put(0,70){(e)} \end{overpic} & 
	\begin{overpic}[type=pdf,ext=.pdf,read=.pdf,width=0.5\linewidth]{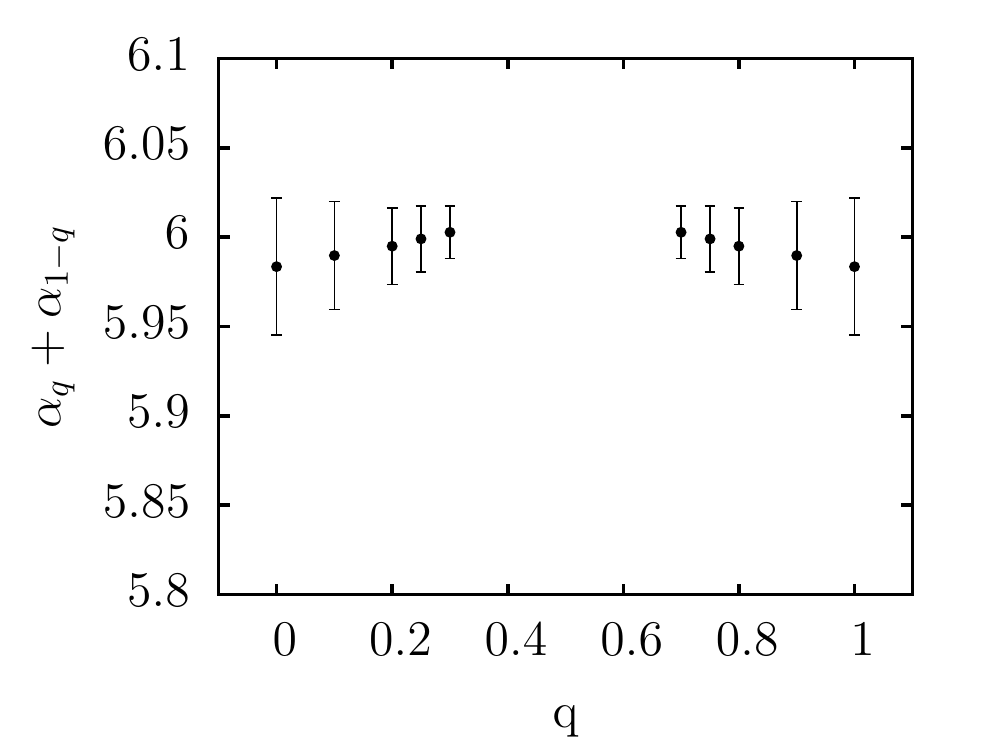}
	\put(0,70){(f)} \end{overpic} \\
	\end{tabular}
	\caption{Symmetry relation of $\Delta_q$ (left column) and $\alpha_q$ (right column) of the 3D quantum percolation model at 
	(a) and (b) $E=0.1$, (c) and (d) $E=0.7$ (at the bottom of the mobility edge), (e) and (f) $E=3.1$. 
	Error bars are $95\%$ confidence levels. Points are naturally symmetric (antisymmetric) for $q=0.5$ for $\alpha_q$ ($\Delta_q$)
	because of the addition (subtraction) of terms corresponding to $q$ and $1-q$.}
	\label{fig:qperc_Deltaalpha_symm_E}	
	\end{center}
\end{figure*}

\begin{figure}
\begin {center}
	\begin{overpic}[type=pdf,ext=.pdf,read=.pdf,width=\linewidth]{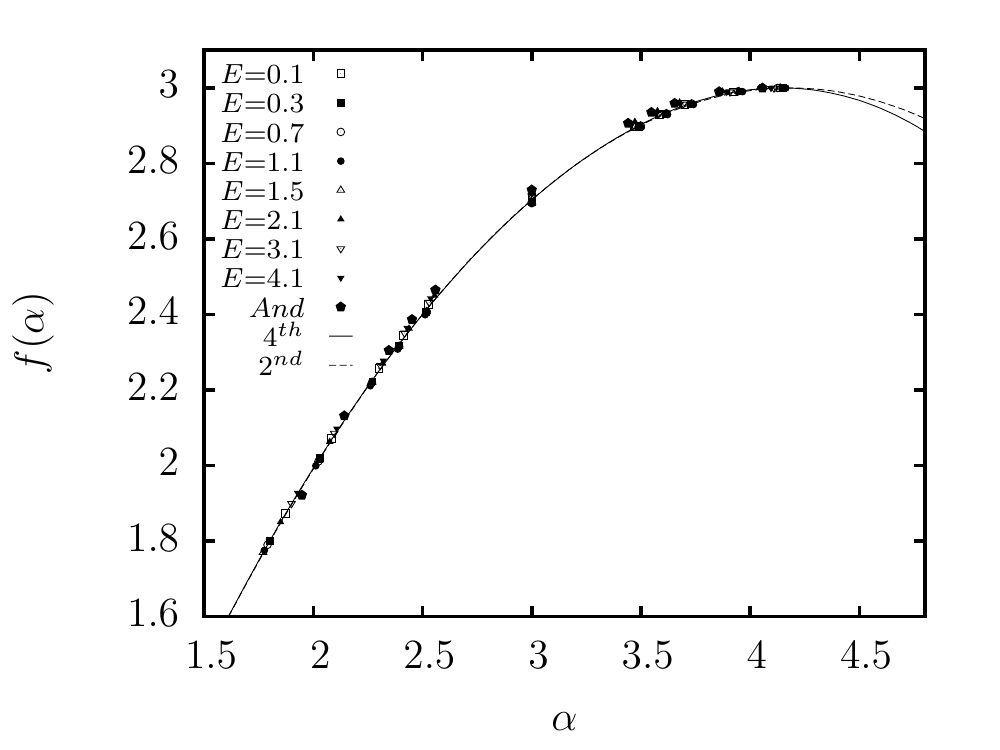}
	\put(50,14){\includegraphics[type=pdf,ext=.pdf,read=.pdf,width=.42\linewidth]{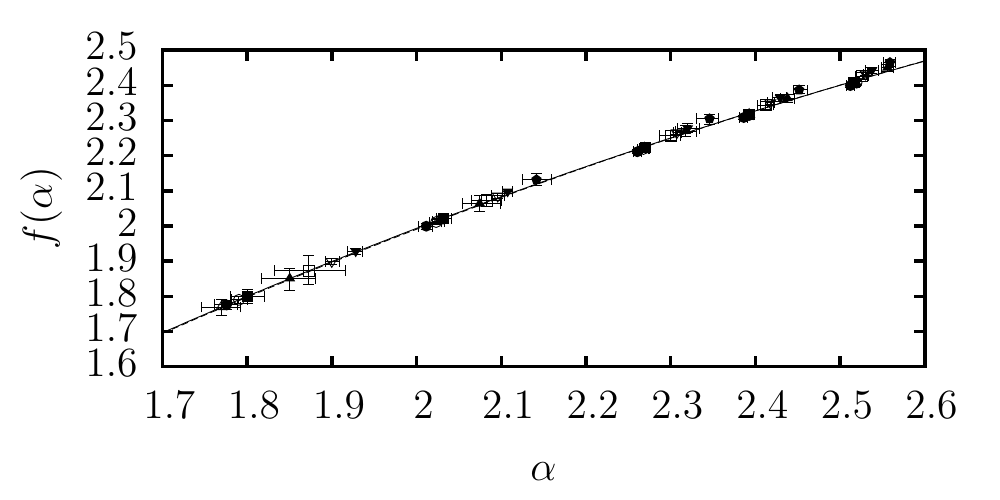}}
	\put(50,34){\includegraphics[type=pdf,ext=.pdf,read=.pdf,width=.42\linewidth]{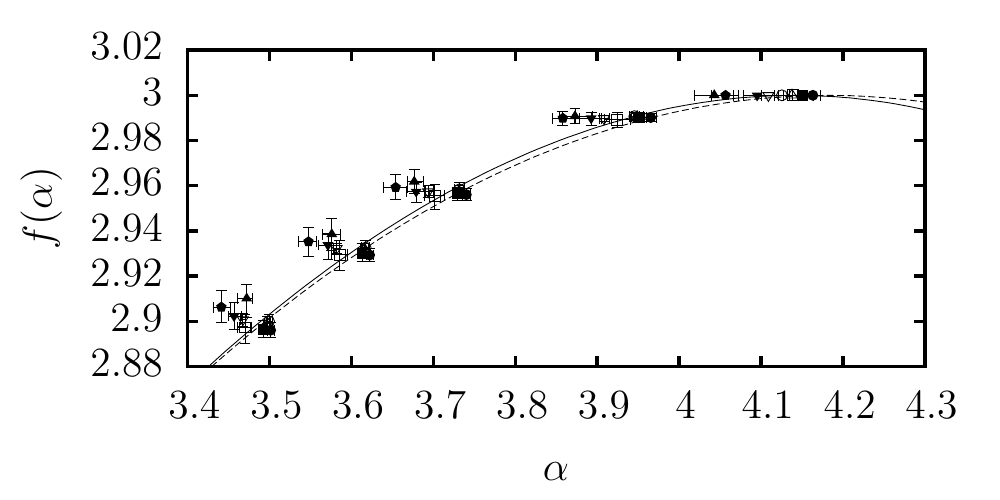}}
	\end{overpic} 
	\caption{$f(\alpha)$ obtained from $D_q$ and $\alpha_q$ computed at different energies, meaning different $p_c^{\scriptscriptstyle Q}$. 
	Pentagons are the results for the Anderson-model, solid line is a $4$th order polynomial, dashed line is a $2$nd order polynomial. Insets are magnified parts 
	of the curve.}
	\label{fig:qperc_falpha}	
	\end{center}
\end{figure}

MFSS provided us the points of the $D_q(E)$ and $\alpha_q(E)$ surface at the investigated energies and $q$ values. By inversion of the mobility edge curve,
$p_c^{\scriptscriptstyle Q}(E)$ one can derive the MFEs as a function of $p_c^{\scriptscriptstyle Q}$ and of $q$, see Fig.~\ref{fig:qperc_Dq_alphaq_shift}. 
Since $D_0=d$, at small $q$ values, i.e. $q\to 0$, the results for $D_q$ are $p_c^{\scriptscriptstyle Q}$-independent, but for larger values of $q$ the $D_q$ starts 
to shift down with decreasing $p_c^{\scriptscriptstyle Q}$, which shows up in the lower right corner of Fig.~\ref{fig:qperc_Dq_alphaq_shift}(a). In  the lower regime of 
Fig.~\ref{fig:qperc_Dq_alphaq_shift}(c) this shift is visibly significant. The same phenomenon can be detected for $\alpha_q$. 
This suggests, that $D_q$ and $\alpha_q$ seem not to behave as  universal quantities.
At relatively larger values of $p_c^{\scriptscriptstyle Q}$, $D_q$ and $\alpha_q$ fulfill the symmetry relation (\ref{eq:multifractals_Deltaalphasymmety}), 
see Fig.~\ref{fig:qperc_Deltaalpha_symm_E} (a), (b), (e) and (f). However, at the bottom of the mobility edge, where $p_c^{\scriptscriptstyle Q}$ 
is smaller, meaning that the lattice is more diluted or more irregular, deviations from the symmetry law seem to be prominent. The $D_q$ and $\alpha_q$ 
values remain the same at small $q$, i.e. when $q\to 0$, but drop down as $q$ increases. 
Resulting in a conclusion, that the symmetry relation, Eq,~(\ref{eq:multifractals_Deltaalphasymmety}), is violated in this regime, 
see for example Fig.~\ref{fig:qperc_Deltaalpha_symm_E} (c) and (d). 
The non-universality of $D_q$ and $\alpha_q$ would automatically imply the non-universality of $\tau_q$, as well. On the other hand with a Legendre-transform for $\tau_q$, 
$f(\alpha)$ can be obtained, describing the scaling of the probability distribution of the wave function amplitudes. This distribution should be universal, 
therefore $f(\alpha)$ should be universal, too. Using Eq.~(\ref{eq:multifractal_falpha_tauq}) and (\ref{eq:multifractal_tauq_Dq_deltaq}) immediately follows:
\begin{equation} 
        f(\alpha_q)=q\alpha_q-D_q(q-1).
\label{eq:GMFEs_f_alpha_D}
\end{equation}
From the $\alpha_q$ and $D_q$ exponents presented in Fig.~\ref{fig:qperc_Dq_alphaq_shift}(a) and (b) we computed the $f(\alpha)$ curve, that is depicted 
in Fig.~\ref{fig:qperc_falpha}. The values from different regimes of the mobility edge seem to form a unique curve, but this is mostly due to the scale on the axis. 
The upper inset of Fig.~\ref{fig:qperc_falpha} shows significant differences between data points at different energies. The approximate shape of the curve is 
a parabola, however, a quartic curve fits the data points slightly better.
According to Eq.~(\ref{eq:GMFEs_f_alpha_D}) $q=1$ corresponds to the fixed point of the $f(\alpha)$ function, $f(\alpha_1)=\alpha_1$. For different values of 
$p_c^{\scriptscriptstyle Q}$  the exponent $\alpha_1$ is not unique, leading to a linear regime of the $f(\alpha)$ function, see the lower inset of 
Fig.~\ref{fig:qperc_falpha}. This makes the whole Legendre-transformation difficult, 
since it needs strict convexity. Conversely an $f(\alpha)$ that is not strictly convex would lead to ill-defined $\tau_q$, $D_q$ and $\alpha_q$, like in our 
case, which contradicts universality again. A possible resolution of this contradiction could be, that our result for the MFEs is just simply not complete, 
perhaps a $p$-dependent phenomenon has not been taken into account affecting the results. Since the problem appeared at the bottom of the mobility edge, 
closest to the classical percolation threshold, one possible candidate for such phenomenon is the existence of an additional length scale, namely the correlation 
length of the classical percolation. In order to test it we added this length scale to the fitting function leading to a 3-variable function with number of fit parameters 
$\sim n_{rel}^3$, but we could not fit so many parameters to our dataset. There is only a small difference between the values of the MFEs for the quantum percolation 
model and for the Anderson model, see Fig.~\ref{fig:qperc_Dq_alphaq_shift}, and the symmetry relation (\ref{eq:multifractals_Deltaalphasymmety}) is almost 
valid within the error bar at the bottom of the mobility edge, too, see Fig.~\ref{fig:qperc_Deltaalpha_symm_E}. Therefore another explanation would be, 
that somehow we underestimated the error bars of the MFEs.
In the $p\to 1$ limit, our exponents seem to be close to their value for the Anderson model, that together with our former claim in Sec.~\ref{sec:fss_qperc}
about their matching universality class corroborate this possibility further. We believe, that there is a unique and universal $D_q$, $\alpha_q$ and $f(\alpha)$ 
curve for the quantum percolation method, and it is identical with the one for the Anderson model, that fulfill the symmetry 
relation (\ref{eq:multifractals_Deltaalphasymmety}). 

As a conclusion the present coherent set of data with a coherent technology in deriving critical exponents fulfill our expectations for larger values of 
$p_c^{\scriptscriptstyle Q}\geq 0.5$ but unfortunately unexpected deviations occur for lower values, i.e. $p_c^{\scriptscriptstyle Q}\leq 0.5$.

\section{Summary}
\label{sec:summ}
In the present work we have numerically investigated the quantum percolation model in 3D. We developed the MFSS method by Rodriguez {\it et. al}~
\cite{Rodriguez11} in order to use it for irregular lattices, or even for graphs in the future. First we tested our method on the well-known Anderson-model, 
however, certain numerical issues forced us to restrict our analysis to the interval $0\leq q \leq 1$, we found $q$-independent results in a good 
agreement with the previous high precision values of Ref~\onlinecite{Rodriguez11}. Then we used our method to the quantum percolation model, 
where we found $q$-independent results again. We numerically determined the mobility edge of the system, confirming previous calculations. 
We also gave an explanation for the behavior of the mobility edge near $E=0$ and at high energy. For the critical exponent we got energy-independent 
values within $95\%$ confidence level. The average of these values is the same as the one for the critical exponent for the 
Anderson model, implying that these models belong to the same universality class. We also determined the MFEs $D_q$ and $\alpha_q$ along the mobility edge, 
and for larger values of $p_c^{\scriptscriptstyle Q}$ we found no significant difference from the Anderson model confirming the statement of the same universality 
class further. In this regime the symmetry relation (\ref{eq:multifractals_Deltaalphasymmety}) is fulfilled. On the other hand in the case of lower 
$p_c^{\scriptscriptstyle Q}$ regime the exponents started to deviate violating universality and (\ref{eq:multifractals_Deltaalphasymmety}), 
probably caused by some unexpected $p$-dependent phenomenon. This behavior deserves further attention. 

\begin{acknowledgments}
   The authors are indebted to dr. A. Stathopoulos for his help setting up the numerical method.
   Financial support from OTKA under grant No. K108676, the Alexander von Humboldt Foundation are gratefully acknowledged.
\end{acknowledgments}


\vfill
\begin{thebibliography}{99}
\bibitem{Anderson}P. W. Anderson, {\it Phys. Rev.} {\bf 109}, 1492 (1958).
\bibitem{EversMirlin}F. Evers, A. D. Mirlin, {\it Rev. Mod. Phys.} {\bf 80}, 1355 (2008) and references therein.
\bibitem{Stauffert} D. Stauffert and J. G. Zabolitzky, {\it J. Phys. A: Math. Gen.} {\bf 19} 3705 (1986).
\bibitem{perc1D} K. Christensen and N. R. Moloney, {\it Complexity and Criticality}, Imperial College Press, (2005).
\bibitem{perc2D} M. E. J. Newman and R. M. Ziff, {\it Phys. Rev. Letters} {\bf 85}, 4104 (2000).
\bibitem{Schubert} G. Schubert, H. Fehske, {\it Lec. Not. Phys.} {\bf 762}, 135 (2009).
\bibitem{Kirkpatrick-Eggarter} S. Kirkpatrick, T. P. Eggarter, {\it Phys. Rev. B} {\bf 6}, 3598 (1972).
\bibitem{Kusy} A. Kusy, A. W. Stadler, G. Haldas, R. Sikora, {\it Physica A} {\bf 241}, 403 (1997).
\bibitem{Soukoulis} C. M. Soukoulis, Q. Li, G. S. Grest, {\it Phys. Rev. B} {\bf 45}, 7724 (1992).
\bibitem{Hoshen-Kopelman} J. Hoshen and R. Kopelman, {\it Phys. Rev. B} {\bf14}, 3438 (1976).
\bibitem{Stathopoulos10} A. Stathopoulos and J. R. McCombs {\it ACM Transaction on Mathematical Software} {\bf 37}, 2, 21:1--21:30 (2010)
\bibitem{Bollhofer08} O. Schenk, M. Bollh\"ofer and R. A. R\"omer, {\it SIAM Review} {\bf 50}, 91 (2008).
\bibitem{Virag14} Ch. Bordenave, A. Sen, B. Vir\'ag, arXiv:1308.3755
\bibitem{Naumis02}G. G. Naumis, Ch. Wang and R. A. Barrio, {\it Phys. Rev. B} {\bf 65}, 134203 (2002).
\bibitem{Rodriguez11} A. Rodriguez, L. J. Vasquez, K. Slevin and R. A. R\"{o}mer, {\it Phys. Rev. B} {\bf84}, 134209 (2011).
\bibitem{janssen}M. Janssen, {\it Fluctuations and localization in mesoscopics electron systems} 
                           (World Scientific Lecture Notes in Physics - Vol. 64, Singapore, 2001); M. Janssen, {\it Phys. Rep.} {\bf 295}, 1 (1998).
\bibitem{cuevas}E. Cuevas and V. E. Kravtsov, {\it Phys. Rev. B} {\bf 76}, 235119 (2007);
\bibitem{Mirlin06}A. D. Mirlin, Y. V. Fyodorov, A. Mildenberger, and F. Evers, {\it Phys. Rev. Lett.} {\bf 97}, 046803 (2006).
\bibitem{milden07}A. Mildenberger and F. Evers, {\it Phys. Rev. B} {\bf 75}, 041303(R) (2007).
\bibitem{evers08}F. Evers, A. Mildenberger, and A. D. Mirlin, {\it phys. stat. sol. b} {\bf 245}, 284 (2008); 
                            F. Evers, A. Mildenberger, and A. D. Mirlin, {\it Phys. Rev. Lett.} {\bf 101}, 116803 (2008)
\bibitem{vasquez08}L. J. Vasquez, A. Rodriguez, and R. A. R\"{o}mer, {\it Phys. Rev. B} {\bf78}, 195106 (2008); 
                                 A. Rodriguez, L. J. Vasquez, and R. A. R\"{o}mer, {\it Phys. Rev. B} {\bf78}, 195107 (2008); 
                                 A. Rodriguez, L. J. Vasquez, and R. A. R\"{o}mer, {\it Phys. Rev. Lett.} {\bf 102}, 106406 (2009).
\bibitem{Travenec} I. Travenec, Int. J. Mod. Phys. B, 22, 5217 (2008).
\bibitem{subra06}A. R. Subramaniam, {\it et al.} {\it Phys. Rev. Lett.} {\bf 96}, 126802 (2006).
\bibitem{faez09}S. Faez, A. Strybulevych, J. H. Page, A. Lagendijk, and B. A. van Tiggelen {\it Phys. Rev. Lett.} {\bf 103}, 155703 (2009).
\bibitem{monthus11}C. Monthus and Th. Garel, {\it J. Stat. Mech.} {\bf 2011}, P05005 (2011); 
                                 see also C. Monthus, B. Berche, and Ch. Chatelain, {\it J. Stat. Mech.} {\bf 2009}, P12002 (2009).
\bibitem{Fehske-Schubert}  G. Schubert, H. Fehske: Quantum and Semi-classical Percolation and Breakdown in Disordered Solids, 
                                            {\it Lecture Notes in Physics} {\bf 762}, (2009), pp 1-28.
\bibitem{Root} L. J. Root, J. D. Bauer, J. L. Skinner, {\it Phys. Rev. B} {\bf 37}, 5518 (1988).
\bibitem{Koslowski} Th. Koslowski, W. von Niessen, {\it Phys. Rev. B} {\bf 44}, 9926 (1991).
\bibitem{Berkovits} R. Berkovits, Y. Avishai, {\it Phys. Rev. B} {\bf 53}, R16125(R) (1996).
\bibitem{Kaneko} A. Kaneko, T. Ohtsuki, {\it J. Phys. Soc. Jap.} {\bf 68}, 1488 (1999).
\end{thebibliography}
\end{document}